\documentclass[aps,prl,twocolumn,superscriptaddress]{revtex4-2}
\usepackage{graphicx}
\usepackage{float}
\usepackage[colorlinks=true,linkcolor=blue,citecolor=red]{hyperref}
\usepackage{siunitx}
\usepackage[justification=justified]{caption}
\usepackage{ragged2e}
\DeclareCaptionLabelFormat{mylabel}{#1#2.\hspace{1.5ex}}
\usepackage{tabularx} 
\usepackage{array}
\usepackage{extarrows}
\usepackage[nameinlink]{cleveref}
\Crefname{equation}{Eq.}{Eqs.}
\Crefformat{equation}{#2Eq.\,(#1)#3}
\Crefname{figure}{Fig.}{Figs.}
\Crefformat{figure}{#2Fig.\,#1#3}
\Crefformat{section}{#2Sec.\,#1#3}   
\newif\ifciteSM
\citeSMfalse

\usepackage{xr}
\makeatletter

\newcommand*{\addFileDependency}[1]{
\typeout{(#1)}
\@addtofilelist{#1}
\IfFileExists{#1}{}{\typeout{No file #1.}}
}\makeatother

\newcommand*{\myexternaldocument}[1]{%
\externaldocument{#1}%
\addFileDependency{#1.tex}%
\addFileDependency{#1.aux}%
}

\myexternaldocument{SupplementaryMaterial}
\begin{document}

\title{
 Oxygen-vacancy quantum spin defects in silicon carbide
}

\author{Yu Chen}
\email{These authors contributed equally to this work.}
\affiliation{Laboratory of Spin Magnetic Resonance, School of Physical Sciences, and Anhui Province Key Laboratory of Scientific Instrument Development and Application, University of Science and Technology of China, Hefei 230026, China}
\affiliation{Hefei National Laboratory, University of Science and Technology of China, Hefei 230088, China}

\author{Qi Zhang*}
\email{zhq2011@ustc.edu.cn}
\email{These authors contributed equally to this work.}
\affiliation{Laboratory of Spin Magnetic Resonance, School of Physical Sciences, and Anhui Province Key Laboratory of Scientific Instrument Development and Application, University of Science and Technology of China, Hefei 230026, China}
\affiliation{School of Biomedical Engineering and Suzhou Institute for Advanced Research, University of Science and Technology of China, Suzhou 215123, China}
\affiliation{Institute of Quantum Sensing, School of Physics, Institute of Fundamental and Transdisciplinary Research, Zhejiang Key Laboratory of R\&D and Application of Cutting-edge Scientific Instruments, Zhejiang University, Hangzhou, 310027, China}

\author{Mingzhe Liu}
\email{These authors contributed equally to this work.}
\affiliation{Laboratory of Spin Magnetic Resonance, School of Physical Sciences, and Anhui Province Key Laboratory of Scientific Instrument Development and Application, University of Science and Technology of China, Hefei 230026, China}

\author{Junda Wu}%
\email{These authors contributed equally to this work.}
\affiliation{Laboratory of Spin Magnetic Resonance, School of Physical Sciences, and Anhui Province Key Laboratory of Scientific Instrument Development and Application, University of Science and Technology of China, Hefei 230026, China}
\affiliation{School of Biomedical Engineering and Suzhou Institute for Advanced Research, University of Science and Technology of China, Suzhou 215123, China}

\author{Jinpeng Liu}%
\affiliation{Laboratory of Spin Magnetic Resonance, School of Physical Sciences, and Anhui Province Key Laboratory of Scientific Instrument Development and Application, University of Science and Technology of China, Hefei 230026, China}
\affiliation{School of Biomedical Engineering and Suzhou Institute for Advanced Research, University of Science and Technology of China, Suzhou 215123, China}

\author{Xin Zhao}
\affiliation{Laboratory of Spin Magnetic Resonance, School of Physical Sciences, and Anhui Province Key Laboratory of Scientific Instrument Development and Application, University of Science and Technology of China, Hefei 230026, China}

\author{Jingyang Zhou}
\affiliation{Laboratory of Spin Magnetic Resonance, School of Physical Sciences, and Anhui Province Key Laboratory of Scientific Instrument Development and Application, University of Science and Technology of China, Hefei 230026, China}

\author{Pei Yu}
\affiliation{Laboratory of Spin Magnetic Resonance, School of Physical Sciences, and Anhui Province Key Laboratory of Scientific Instrument Development and Application, University of Science and Technology of China, Hefei 230026, China}

\author{Shaochun Lin}
\affiliation{Laboratory of Spin Magnetic Resonance, School of Physical Sciences, and Anhui Province Key Laboratory of Scientific Instrument Development and Application, University of Science and Technology of China, Hefei 230026, China}

\author{Yuanhong Teng}
\affiliation{Laboratory of Spin Magnetic Resonance, School of Physical Sciences, and Anhui Province Key Laboratory of Scientific Instrument Development and Application, University of Science and Technology of China, Hefei 230026, China}

\author{Wancheng Yu}
\affiliation{State Key Laboratory of Crystal Materials, Institute of Novel Semiconductors, Shandong University, Jinan 250100, China}

\author{Ya Wang}
\affiliation{Laboratory of Spin Magnetic Resonance, School of Physical Sciences, and Anhui Province Key Laboratory of Scientific Instrument Development and Application, University of Science and Technology of China, Hefei 230026, China}
\affiliation{Hefei National Laboratory, University of Science and Technology of China, Hefei 230088, China}
\affiliation{Hefei National Research Center for Physical Sciences at the Microscale, University of Science and Technology of China, Hefei 230026, China}

\author{Chang-Kui Duan}
\affiliation{Laboratory of Spin Magnetic Resonance, School of Physical Sciences, and Anhui Province Key Laboratory of Scientific Instrument Development and Application, University of Science and Technology of China, Hefei 230026, China}
\affiliation{Hefei National Laboratory, University of Science and Technology of China, Hefei 230088, China}
\affiliation{Hefei National Research Center for Physical Sciences at the Microscale, University of Science and Technology of China, Hefei 230026, China}

\author{Fazhan Shi}
\email{fzshi@ustc.edu.cn}
\affiliation{Laboratory of Spin Magnetic Resonance, School of Physical Sciences, and Anhui Province Key Laboratory of Scientific Instrument Development and Application, University of Science and Technology of China, Hefei 230026, China}
\affiliation{Hefei National Laboratory, University of Science and Technology of China, Hefei 230088, China}
\affiliation{School of Biomedical Engineering and Suzhou Institute for Advanced Research, University of Science and Technology of China, Suzhou 215123, China}
\affiliation{Hefei National Research Center for Physical Sciences at the Microscale, University of Science and Technology of China, Hefei 230026, China}

% Abstract should be written in the present tense and impersonal style (i.e., avoid we), and be at most 200 words long
\begin{abstract}
Optically addressable spin defects in wide-bandgap semiconductors are promising building blocks for quantum sensing and quantum networks. Establishing their atomic structure is essential for understanding functionality and enabling controlled engineering. 
In 4H-SiC, the PL5 and PL6 centers have long been recognized for their exceptional charge stability and room-temperature optically detected magnetic resonance (ODMR) performance, but their structural origin has remained elusive for over a decade. Here, we provide direct evidence for their oxygen-vacancy ($\mathrm{O_C V_{Si}}$) origins through a combined chemical and isotopic control strategy. Under oxygen ion implantation, we observe over tenfold enhancement in the yield of PL5 and PL6 compared to nitrogen ion implantation. Furthermore, implantation with $^{17}\text{O}$ ions produces PL5 and PL6 defects that exhibit a characteristic six‑fold $^{17}\text{O}$ hyperfine splitting in their ODMR spectra.
These results affirm PL6 as the $\mathrm{O_C V_{Si}}$ defect in the $hh$ configuration. For PL5, the oxygen-related evidence, together with \textit{ab initio} calculations and additional measurements of the zero-field splitting and hyperfine structure, establishes it as the $\mathrm{O_C V_{Si}}$ defect in the $kh$ configuration.
This unambiguous structural identification, achieved through materials-level chemical control, provides the microscopic foundation for deterministic engineering of these defects, paving the way for scalable photonic devices and high-sensitivity ensemble quantum sensors based on oxygen-vacancy centers.
\end{abstract}
\maketitle

\begin{figure*}
\centering 
\includegraphics{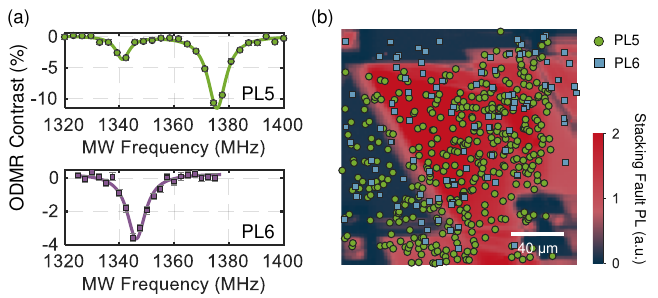}
\caption{(a) Zero-field ODMR spectra of PL5 and PL6. (b) Overlay of the stacking-fault photoluminescence map with the locations of individually identified PL5 (circles) and PL6 (squares) centers. The stacking-fault PL is obtained by integrating the emission intensity over the 410–430 nm spectral range under 325 nm laser excitation.}
\label{Figure1}
\end{figure*}

\section{Introduction}

Optically addressable spin defects in solids have emerged as a versatile platform for quantum technologies, enabling applications in quantum sensing, quantum communication, and distributed quantum networks \cite{2017-QuantumSensing-Degen-Rev.Mod.Phys.,2018-MaterialPlatformsSpinbased-Atature-NatRevMater}. Their localized electron spins provide long-lived quantum states, while optical transitions enable efficient initialization, control, and readout, as well as spin–photon interfaces. A central challenge in this field is identifying the atomic structures of these defects, as the microscopic configuration of impurities and vacancies governs their electronic structure, spin interactions, optical transitions, and charge stability \cite{wolfowiczQuantumGuidelinesSolidstate2021}. Determining defect structures is therefore essential both for understanding their physical properties and for enabling deterministic defect engineering in scalable materials platforms.

Silicon carbide (SiC) has recently emerged as a promising host for such quantum defects owing to its wafer-scale growth, compatibility with semiconductor fabrication, and the presence of spin defects with long coherence times, such as divacancies (labeled as PL1–PL4) and silicon vacancies \cite{2020-DevelopingSiliconCarbide-Son-AppliedPhysicsLetters}. 
Among the optically active defects in 4H-SiC, the unresolved centers labeled PL5 and PL6 are particularly attractive. Their room-temperature spin properties are comparable to those of the nitrogen–vacancy (NV) center in diamond, while their infrared emission and compatibility with semiconductor processing provide additional advantages \cite{liRoomtemperatureCoherentManipulation2022,koehlRoomTemperatureCoherent2011}. In our measurements, even without photonic-structure enhancement, single PL6 and PL5 centers exhibit fluorescence rates of 460 kcps and 250 kcps, respectively (\ifciteSM\Cref{figS: PL5-6 Saturation Count} \else Fig.\,S1 \fi in \cite{SupplementalMaterial}), highlighting their promise for quantum sensing applications. Despite these favorable properties, the generation yield of PL5 and PL6 remains significantly lower than that of other defects in SiC \cite{koehlRoomTemperatureCoherent2011}, and theoretical guidance for improving their formation efficiency has been limited by the lack of confirmed microscopic models.

Over the past decade, several microscopic models have been proposed to explain the origins of PL5 and PL6, including divacancies
stabilized near stacking faults and oxygen–vacancy complexes \cite{ivadyStabilizationPointdefectSpin2019,2025-OriginUnidentifiedColor-Bai-Phys.Rev.B,2025-IdentifyingPL6Center-Zhao-Phys.Rev.Mater.a}. However, because of the intrinsic accuracy limits of first-principles calculations, different defect configurations can produce very similar predicted spectroscopic signatures, and thus theoretical predictions alone cannot unambiguously determine the defect structure. As a result, the atomic structures of PL5 and PL6 remain unresolved and are still unambiguously referred as divacancy-related or unidentified defects in the literature \cite{2025-StrainEnhancedSpinReadout-Hu-Phys.Rev.Lett.,2025-NoninvasiveBioinertRoomtemperature-Li-Nat.Mater.,2025-ProbingNoiseSpectrum-Zhao-Phys.Rev.Appl.}.

Here we resolve this longstanding question by providing direct experimental evidence that excludes the involvement of stacking faults and demonstrates oxygen incorporation in the structures of PL5 and PL6. Using correlative imaging with single-defect spatial resolution, we show that these defects form independently of stacking faults.  By implementing a combined chemical and isotopic control strategy, we further confirm oxygen incorporation into their structures. First, by replacing conventional nitrogen ion implantation with oxygen ion implantation, we observe a substantial increase in defect generation efficiency, with the yields of PL5 and PL6 enhanced by more than 11-fold and 23-fold, respectively, indicating that oxygen plays a catalytic role in their formation. Second, implantation with \(^{17}\text{O}^{+}\) ions produces a characteristic six-fold \(^{17}\text{O}\) hyperfine splitting in the ODMR spectra of both defects, providing direct spectroscopic evidence of oxygen incorporation.

Combining these experimental observations with first-principles calculations \cite{2025-OriginUnidentifiedColor-Bai-Phys.Rev.B,2025-IdentifyingPL6Center-Zhao-Phys.Rev.Mater.a,2023-OxygenvacancyDefect4HSiC-Kobayashi-JournalofAppliedPhysics}, we assign PL6 to the \(\mathrm{O_C V_{Si}}\)(\(hh\)) configuration. For PL5, however, measurements of the full orientation of the zero-field-splitting tensor (\(D\), \(E\)) and hyperfine couplings, together with first-principles calculations, identify the defect as the \(\mathrm{O_C V_{Si}}\)(\(kh\)) configuration rather than the previously proposed \(hk\) assignment \cite{2025-OriginUnidentifiedColor-Bai-Phys.Rev.B}. The unambiguous identification of the atomic structures of PL5 and PL6 not only resolves a long-standing materials puzzle but also provides a foundation for rational defect engineering in 4H-SiC. Building on these structural insights, we discuss at the end of this article how the atomic configuration informs potential strategies to improve defect formation and deterministic placement, highlighting avenues for designing high-performance quantum devices.

\section{PL5/PL6 Form Independently of Stacking Faults }

Previous studies based on ensemble-averaged measurements have suggested that PL5 and PL6 spin defects are confined near the surface of silicon carbide and attributed this spatial distribution to a correlation with near-surface stacking faults~\cite{ivadyStabilizationPointdefectSpin2019,shafizadehEvolutionOpticallyDetected2025}. However, the limited spatial resolution of these ensemble techniques has prevented a definitive conclusion. In this work, we overcome this limitation by probing the correlation at the single-defect level, which reveals a different picture: PL5 and PL6 are found to form independently of stacking faults. We first mapped the locations of individual PL5 and PL6 defects using single-spin ODMR spectroscopy; representative ODMR spectra from our sample are shown in \Cref{Figure1}(a). We then mapped the spatial distribution of stacking faults (SFs) within the same region using a photoluminescence (PL) based method described in Ref.~\cite{fengCharacterizationStackingFaults2008}. Under 325-nm laser illumination, the intrinsic 4H-SiC PL spectrum exhibits a band-edge emission peak at 386,nm (3.21,eV). The presence of an SF locally reduces the bandgap, causing a redshift in the PL spectrum~\cite{ivadyStabilizationPointdefectSpin2019,fengCharacterizationStackingFaults2008} that enables its selective detection. We collected PL spectra over a \SI{180}{\micro\meter}$\times$\SI{180}{\micro\meter} region spanning wavelengths from 400\,nm to 520\,nm. Within this region, only PL peaks at 420\,nm (2.95\,eV) were observed, corresponding to two types of SFs: single Shockley SFs and intrinsic Frank SFs~\cite{fengCharacterizationStackingFaults2008} (see \ifciteSM\Cref{figS: SF_PL_Image} \else Fig.\,S3 \fi in Supplementary Information~\cite{SupplementalMaterial}). \Cref{Figure1}(b) shows the superimposed locations of PL5, PL6, and the stacking faults after spatial alignment. The results clearly demonstrate that PL5 and PL6 can form far away from stacking faults, indicating that these defects are not correlated with stacking fault structures.

\section{Oxygen Incorporation in PL5/PL6 Atomic Structures}
\begin{figure}
\centering 
\includegraphics{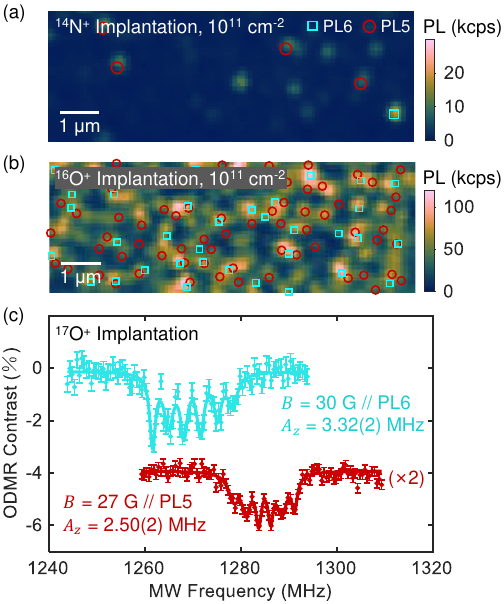}
\caption{
\textbf{(a)} Photoluminescence map of a nitrogen-ion-implanted sample. Implantation was carried out at 15\,keV with a dose of $10^{11}\,\mathrm{cm^{-2}}$, followed by annealing at $1050\,^\circ\mathrm{C}$ for 30\,minutes.
\textbf{(b)} PL map of an oxygen-ion-implanted sample under identical implantation and annealing conditions.
In both (a) and (b), individual PL5 and PL6 centers identified via ODMR are marked with red circles and blue squares, respectively (see Supplementary Information \ifciteSM\Cref{secS: map-O-implanted} \else Sec.\,S2 \fi \cite{SupplementalMaterial}).
\textbf{(c)} ODMR spectra of PL5 (vertically scaled $\times 2$) and PL6 measured in a \(\mathrm{^{17}O^+}\)-implanted sample. The magnetic field is aligned with the defect axis at 27\,G for PL5 and 30\,G for PL6. Solid lines represent fits using six Lorentzian profiles.
}
\label{Figure2}
\end{figure}

To determine whether oxygen is involved in PL5 and PL6, we employed a combined chemical and isotopic control strategy. We directly compared samples implanted with oxygen ions against those implanted with nitrogen ions. Since the intrinsic oxygen concentration in as-grown 4H-SiC CVD epilayers is typically low ($<10^{12}\,\mathrm{cm^{-3}}$) \cite{1999-OxygenSiliconCarbide-Dalibor-MaterialsScienceandEngineering:B,2001-OxygenRelatedDefectCenters-Klettke-Mater.Sci.Forum}, ion implantation provides a controlled means to significantly increase the local oxygen density. If PL5/PL6 indeed arises from an OV complex, a remarkable enhancement in its generation efficiency should be observed under oxygen ion implantation.
\Cref{Figure2}(a--b) shows PL mapping results for samples implanted with oxygen and nitrogen ions under identical conditions (15\,keV, $10^{11}\,\mathrm{cm^{-2}}$, followed by annealing at $1050\,^\circ\mathrm{C}$ for 30\,minutes). The overall PL intensity is notably higher in the oxygen-ion-implanted region. Using ODMR technique to identify individual defects (see Supplementary Information \ifciteSM\Cref{secS: map-O-implanted} \else Sec.\,S2 \fi \cite{SupplementalMaterial}), we located emission spots corresponding to PL5 (circled) and PL6 (squared) in the maps.
Statistical analysis over extended areas (see Supplementary Information \ifciteSM\Cref{secS: map-O-implanted} \else Sec.\,S2 \fi \cite{SupplementalMaterial}) reveals that the generation efficiency of PL5 and PL6 centers increases by factors exceeding 11 and 23, respectively, in oxygen-ion-implanted samples relative to nitrogen-ion-implanted ones. This order-of-magnitude enhancement strongly suggests that oxygen is integral to the formation of these defects.

To further confirm the incorporation of oxygen into the PL5/PL6 atomic structure, we performed ODMR measurements on a sample implanted with the stable isotope \(\mathrm{^{17}O}\) (nuclear spin \(I = 5/2\)) ions. As shown in \Cref{Figure2}(c), the ODMR spectra of PL5 and PL6 exhibit a characteristic six-fold hyperfine splitting when the magnetic field is aligned with the defect axis. The spectra are well described by six Lorentzian peaks, from which we extract the axial hyperfine coupling \(A_z\). The fitted values are \(2.50(2)\,\mathrm{MHz}\) for PL5 and \(3.32(2)\,\mathrm{MHz}\) for PL6, similar to the calculated \(\mathrm{^{17}O}\) hyperfine couplings for the \(\mathrm{O_C V_{Si}}\) defect \cite{2023-OxygenvacancyDefect4HSiC-Kobayashi-JournalofAppliedPhysics}. This isotopic signature provides direct evidence that oxygen is a constituent of the PL5/PL6 defect complex.

\section{Identifying PL5 as the $\mathrm{O_C V_{Si}}$($kh$) Configuration}

Having established that oxygen is integral to both PL5 and PL6, whereas stacking faults are excluded, the oxygen–vacancy complex $\mathrm{O_C V_{Si}}$ emerges as the most plausible structural model.
In 4H-SiC,  the $\mathrm{O_C V_{Si}}$ defect can adopt four distinct site configurations ($hh$, $kk$, $hk$, $kh$), determined by whether the oxygen atom and the silicon vacancy occupy sites in hexagonal ($h$) or quasi-cubic ($k$) crystal environments.
For PL6, the unambiguous \(^{17}\text{O}\) hyperfine signature obtained in this work, along with previous measurements and theoretical analyses of its zero-field splitting and hyperfine structure \cite{2025-OriginUnidentifiedColor-Bai-Phys.Rev.B,2025-IdentifyingPL6Center-Zhao-Phys.Rev.Mater.a}, is able to convincingly identifies it as the $\mathrm{O_C V_{Si}}(hh)$ configuration.
In contrast, key spectral features of PL5, particularly its hyperfine coupling patterns and the transverse zero-field splitting $E$, have not been sufficiently characterized or compared with $\mathrm{O_C V_{Si}}$ calculations to determine its structure definitively. In following part, we complete the characterization of PL5 through comprehensive zero-field splitting and hyperfine structure measurements.

\begin{figure}
\centering 
\includegraphics{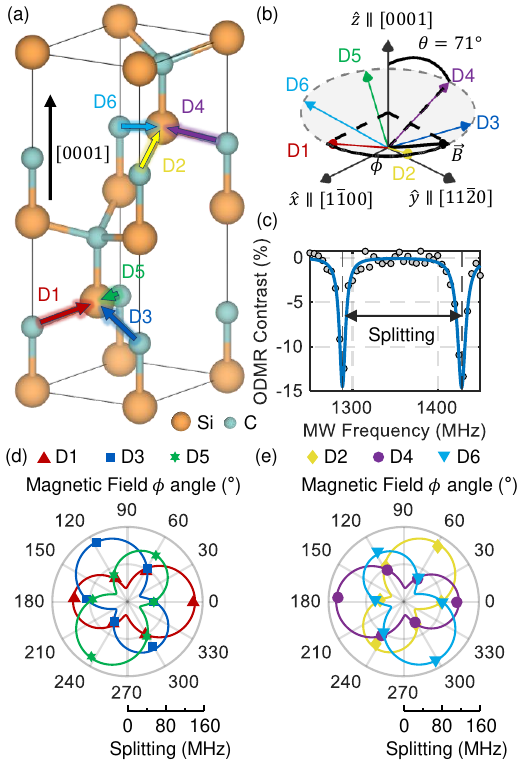}
\caption{ (a) Schematic of six PL5 directions D1--D6 in 4H-SiC lattice. (b) Schematic of six PL5 directions, coordinate axes, and the magnetic field for measurements in (d-e). The magnitude of the magnetic field was set to 25 Gauss. (c) ODMR spectrum of a D1 direction PL5 under $\phi=30^{\circ}$. (d)(e) Polar plots of ODMR splittings of PL5 of different orientations as functions of the magnetic orientation $\phi$. }
\label{Figure3}
\end{figure}

The zero-field splitting parameters of PL5, $D$ and $E$, were previously measured as 1373 MHz and 16.5 MHz at low temperature \cite{falkPolytypeControlSpin2013}.
Here, we further resolve their orientations, which provide important constraints on the atomic structure. We found that PL5 centers are equally distributed among the six orientations of the basal C-Si bonds in 4H-SiC [\Cref{Figure3}(a)]. This observation is further confirmed by rotational magnetic-field measurements. By rotating the magnetic field on a conical surface at a 71\,$^\circ$ angle to the c-axis and measuring the splitting of PL5 centers [\Cref{Figure3}(b--c)], we found there are six distinct angular dependences as depicted in \Cref{Figure3}(d--e). Each dependence corresponds to one direction of PL5, revealing six defect orientations (D1--D6) of PL5. Previous studies reported only three orientations for basal divacancies, mostly because the applied fields (microwave, strain, and excitation laser) were confined within the c-plane in those studies \cite{falkElectricallyMechanicallyTunable2014,koehlRoomTemperatureCoherent2011,liRoomtemperatureCoherentManipulation2022}.

\begin{figure}
\includegraphics{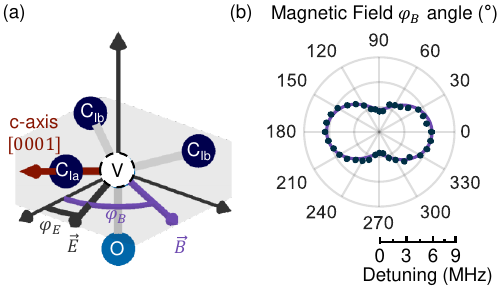}
\centering 
\caption{(a) Schematic of PL5, coordinate axes and the magnetic field. The magnitude of the magnetic field was set to 25 Gauss. The coordinate axes are defined such that the direction of axially symmetric splitting is selected as the z-axis. The coordinate system is chosen such that the c-axis lies within the xz plane. (b) Polar plot of the measured detuning as a function of the magnetic orientation $\varphi_{B}$. The detuning is set from $f_0= 1375.3\ \text{MHz}$. Fitting using \Cref{eq:omega_pm} yields $\varphi_E = 182.2\pm 3.5 ^{\circ}$.}
\label{Figure4} 
\end{figure}

The coordinate system for measuring the direction of $E$ is defined such that the direction of $D$ aligns with the z-axis, while the crystal c-axis lies within the xz plane [\Cref{Figure4}(a)]. A magnetic field of 25 Gauss was rotated around the z-axis to measure the azimuth angle $\varphi_E$ of $E$ \cite{doldeElectricfieldSensingUsing2011a}. The two magnetic dipole transition frequencies measured on the ODMR spectrum of PL5 are given by \cite{doldeElectricfieldSensingUsing2011a,dohertyTheoryGroundstateSpin2012}
\begin{equation}
  f_{\pm}= D+3 \eta \pm (E^2-2 \eta \cos (2\varphi_B+\varphi_E) + \eta ^2)^{\frac{1}{2}}
\label{eq:omega_pm}
\end{equation}
where \(\varphi_B\) is the azimuth angle of magnetic field \(B\) and \(\eta = (\gamma B)^2/2D\). \Cref{Figure4}(b) shows the measured \(f_+\) as a function of \(\varphi_B\), and fitting with \Cref{eq:omega_pm} yields $\varphi_E = 182.2\pm 3.5 ^{\circ}$, indicating that the $E$ direction lies within the defect symmetry plane.

\begin{figure}
\centering 
\includegraphics{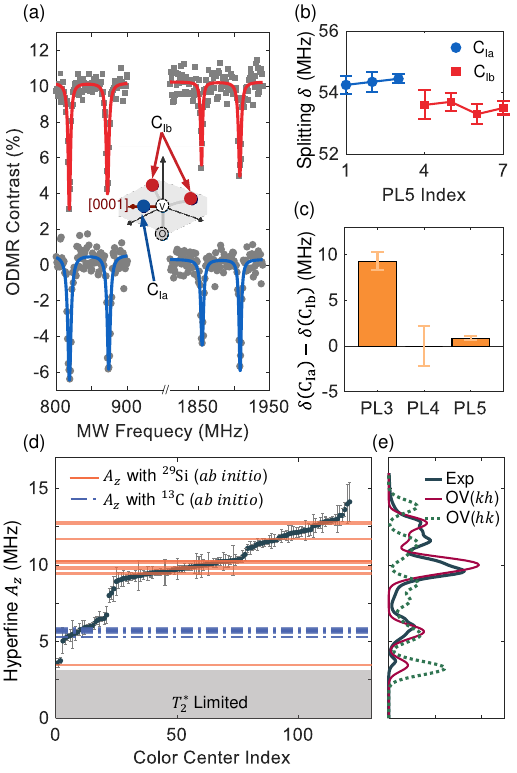}
\caption{(a) ODMR spectra of PL5 coupled to a $^{13}\text{C}_{\text{I}}$ nuclear spin adjacent to the $\text{V}_{\text{Si}}$ in PL5. Blue(Red) line represents the fit for the data of PL5 coupled with $^{13}\text{C}_{\text{Ia}} (^{13}\text{C}_{\text{Ib}})$. All spectra were acquired under an applied magnetic field of 185 Gauss. (b) Hyperfine splitting $\delta$ induced by the interaction with $^{13}\text{C}_{\text{Ia}} $(blue) and $^{13}\text{C}_{\text{Ib}}$(red), as derived from different PL5. (c) The splitting difference $\delta(\text{C}_{\text{Ia}})-\delta(\text{C}_{\text{Ib}})$ for PL3\textendash PL5. Experimental data for PL3 and PL4 are reproduced from \cite{sonDivacancy4HSiC2006}. (d) Measured hyperfine constant $A_z$ for different PL5 centers under an applied magnetic field of $\gamma B_z = 39.6 \pm 1\ \text{MHz}$ \cite{SupplementalMaterial}. The color centers are sorted by $A_z$ for clarity. Solid (dashed) lines represent $\mathrm{O_C V_{Si}}(kh)$ first-principles results of couplings to $^{29}\text{Si}$ ($^{13}\text{C}$) spins at different lattice locations. The hyperfine couplings less than 3\,MHz cannot be detected due to the ODMR broadening with $T_2^*$, marked by the shadow area. (e) Experimental (black solid curve) versus simulated $\mathrm{O_C V_{Si}}(kh)$ (red solid curve) and $\mathrm{O_C V_{Si}}(hk)$ (green dashed curve) $A_z$ distributions. The experimental $A_z$ distribution is generated by Gaussian broadening by broadening each data point in (d) as a Gaussian peak and then summing. The error bar of each point serves as the standard deviation for each experimental Gaussian peak. The simulated $A_z$ distribution is given by the $\mathrm{O_C V_{Si}}(kh)$ and $\mathrm{O_C V_{Si}}(hk)$ first-principles results, with broadening of the average of the experimental error bars.}
\label{Figure5} 
\end{figure}

The hyperfine coupling measurements of PL5 were performed in two stages. First, under an aligned magnetic field, we recorded the ODMR spectra of PL5 centers coupled to a $^{13}\text{C}_{\text{Ia}} $ or $^{13}\text{C}_{\text{Ib}}$ nuclear spin, which exhibit the strongest hyperfine interaction with the defect [\Cref{Figure5}(a)]. Slight differences in hyperfine splitting were observed between the $|0\rangle \leftrightarrow |+1\rangle$ and $|0\rangle \leftrightarrow |-1\rangle$ transitions (see Supplementary Information \ifciteSM\Cref{tabS: Nearest_C_Coupling}\else Table\,S4\fi). We define the average of these two splittings as the characteristic hyperfine splitting $\delta$. We measured $\delta$ for three PL5 centers coupled to $^{13}\text{C}_{\text{Ia}}$ and four to $^{13}\text{C}_{\text{Ib}}$ [\Cref{Figure5}(b)]. The $^{13}\text{C}_{\text{Ia}}$-coupled centers exhibited splittings larger by 0.88 MHz on average compared to their $ ^{13}\text{C}_{\text{Ib}}$ counterparts. The splitting differences $\delta(\mathrm{C_{Ia}})-\delta(\mathrm{C_{Ib}})$ for PL3–PL5 are summarized in \Cref{Figure5}(c) for later comparison with theoretical results.
In the second stage, using a lower magnetic field generated by Helmholtz coils, we measured 128 PL5 centers that showed strong coupling to nuclear spins ($^{29}\text{Si}$, $^{13}\text{C}$) other than $\mathrm{C_I}$ (see Supplementary Information \ifciteSM\Cref{secS:Measurement of Hyperfine Couplings in PL5} \else Sec.\,S4 \fi for details \cite{SupplementalMaterial}). The corresponding axial hyperfine couplings $A_z$, calculated from these spectra, are plotted in ascending order for clarity in \Cref{Figure5}(d). To better visualize the distribution, each measured $A_z$ value was represented by a Gaussian peak (with its value as the center and the uncertainty as the variance) ; the sum of all such peaks is shown in \Cref{Figure5}(e).

Experimental measurements of the zero-field splitting parameters $D$ and $E$, together with the hyperfine couplings, enable direct comparison with theoretical predictions. The measured $D$ orientation restricts the possible PL5 structures to a basal-plane oxygen-vacancy complex, i.e., either the $hk$ or $kh$ configuration of $\mathrm{O_C V_{Si}}$. To compare with PL5 and to assess the accuracy of our computational approach, we performed first-principles calculations on basal neutral oxygen-substituted silicon vacancies as well as on basal divacancies. Computational details are provided in the Supplementary Information \cite{SupplementalMaterial}, and the main results are summarized in \Cref{tab: DFT-T} and \Cref{Figure5}(e).

\begin{table*}
\centering
\caption{ Comparison of selected first-principles results with experimental properties of PL3--PL5. Experimental values for ZPL, $D$, and $E$ at low temperature are reported in \cite{falkPolytypeControlSpin2013,falkElectricallyMechanicallyTunable2014}. For PL5, $E_x = E\cos{\varphi_E} $ and $E_y = E\sin{\varphi_E} $, where $\varphi_E$ is measured in this paper. $\delta(\text{C}_{\text{Ia}})$ and $\delta(\text{C}_{\text{Ib}})$ are hyperfine splittings due to the nearest $^{13} \text{C}$ nuclei $\text{C}_{\text{Ia}}$ and $\text{C}_{\text{Ib}}$. The experimental splitting differences $\delta(\text{C}_{\text{Ia}})-\delta(\text{C}_{\text{Ib}})$ for PL3--PL5 are obtained from \Cref{Figure5}(c).
}
\label{tab: DFT-T}
\renewcommand{\arraystretch}{1.5}
\setlength{\tabcolsep}{7pt}
\begin{tabular}{lccccc}
\hline \hline
Label&ZPL&$D$&$E_x$&$|E_y|$& $\delta(\text{C}_{\text{Ia}})-\delta(\text{C}_{\text{Ib}})$\\
&(eV)&(MHz)&(MHz)&(MHz)& (MHz)\\
\hline
(Exp.) PL5& 1.189 & 1373& -16.5& $<$0.5& 0.88\\
 (Cal.) $\mathrm{O_{C}V_{Si}}(kh)$ & 1.12 & 1552 & -47.0 & 7.7 &0.89\\
 (Cal.) $\mathrm{O_{C}V_{Si}}(hk)$ & 1.06 & 1440 & -96.5 & 7.8 &2.32
\\
\hline
(Exp.) PL4 & 1.118& 1334& \multicolumn{2}{c}{$(E_x^2+E_y^2)^{1/2}=18.7$} & <3\\
(Cal.) \(\text{V}_{\text{C}}\text{V}_{\text{Si}}\)($kh$)& 1.12 & 1465 & -33.9& 0.0 & 1.72\\
 (Exp.) PL3& 1.119 & 1222& \multicolumn{2}{c}{$(E_x^2+E_y^2)^{1/2}=82$} & 9.3\\
(Cal.) \(\text{V}_{\text{C}}\text{V}_{\text{Si}}\)($hk$)& 1.08 & 1402 & -63.9& 0.0 & 4.57\\
 \hline \hline
\end{tabular}
\end{table*}

Comparison between calculated and experimentally established properties of the identified divacancies PL3 [($\mathrm{V_C V_{Si}}(hk)$] and PL4 [$\mathrm{V_C V_{Si}}(kh)$] indicates that our calculations overestimate $D$ by no more than 200\,MHz, while the deviations in $E$ and the zero-phonon line (ZPL) energy do not exceed 30\,MHz and 0.05\,eV, respectively (\Cref{tab: DFT-T}). Within these error margins, the calculated properties of $\mathrm{O_C V_{Si}}(kh)$ are consistent with the experimental data for PL5 (\Cref{tab: DFT-T}). Moreover, the first-principles results for $\mathrm{O_C V_{Si}}(kh)$ show better agreement with PL5 than those for $\mathrm{O_C V_{Si}}(hk)$, both in terms of the $E$ parameter (\Cref{tab: DFT-T}) and the statistics of hyperfine couplings (\Cref{Figure5}(e)). These findings support assigning the $kh$ configuration to PL5, thereby revising the previous $hk$ assignment \cite{2025-OriginUnidentifiedColor-Bai-Phys.Rev.B}.

\section{Conclusion and Discussion}

This work resolves the long-standing puzzle regarding the structural origin of the PL5 and PL6 centers in 4H-SiC. By combining isotope-controlled ion implantation with spectroscopic analysis, we establish direct oxygen–vacancy origins for these defects. The characteristic six-fold hyperfine structure observed in samples implanted with \(\mathrm{^{17}O^+}\) provides unambiguous evidence. Supported by first-principles calculations, we assign PL5 to the \(\mathrm{O_C V_{Si}}\) defect in the \textit{kh} configuration and confirm PL6 as its \textit{hh} counterpart.

The identification of the \(\mathrm{O_C V_{Si}}\) structure suggests formation kinetics for these promising quantum defects is analogous to those of the NV center in diamond, both governed by the pairing of an extrinsic atom with a vacancy. This finding overturns the prevailing assumption that PL5 and PL6 are inherently surface-confined defects \cite{ivadyStabilizationPointdefectSpin2019, shafizadehEvolutionOpticallyDetected2025}, and reestablishes electron irradiation combined with oxygen incorporation as a viable route for generating high-quality, three-dimensional PL5/PL6 ensemble quantum sensors \cite{sonModifiedDivacancies4HSiC2022}. The spatial control strategies developed for NV centers in diamond may also be adapted for \(\mathrm{O_C V_{Si}}\) \cite{2022-SelfalignedPatterningTechnique-Wang-Sci.Adv.}, enabling deterministic placement of PL5 and PL6 centers within photonic nanostructures---a key step toward realizing efficient spin–photon interfaces.

With the defect configuration now established, we anticipate future studies on the formation energetics of PL5/PL6, similar to those conducted for divacancies and other spin defects in SiC \cite{leeStabilityMolecularPathways2021, zhangEngineeringFormationSpindefects2023a}. A deeper understanding of the thermodynamic and charge-state conditions that maximize formation probability will guide the optimization of doping protocols, implantation doses, annealing temperatures, and Fermi-level control. Experimentally, our oxygen-implantation approach already enhances the generation yield by more than an order of magnitude. Further refinement assisted by \textit{ab initio} studies may suppress competing defects and elevate ensemble ODMR contrast from the typical sub-percent level toward the intrinsic single-defect contrast exceeding 20\% \cite{falkPolytypeControlSpin2013, liRoomtemperatureCoherentManipulation2022}, which will be a crucial advancement for ensemble-based sensing applications.

In summary, the conclusive identification of PL5 and PL6 as oxygen-vacancy centers establishes a crucial materials-level understanding. 
This knowledge forms the foundation for a rational defect-engineering framework in 4H-SiC, paving the way toward the realization of high-performance, CMOS-compatible quantum sensors and scalable quantum networks based on these versatile spin defects.

% Acknowledgements
\begin{acknowledgments}
This work was supported by the National Natural Science Foundation of China (Grant Nos.\,T2125011, 12174377), the CAS (Grant No.\,YSBR-068), Quantum Science and Technology-National Science and Technology Major Project (Grant Nos.\,2021ZD0302200, 2021ZD0303204), New Cornerstone Science Foundation through the XPLORER PRIZE, "Pioneer" and "Leading Goose" R\&D Program of Zhejiang (Grant No.\,2025C01041)
and the Fundamental Research Funds for the Central Universities (Grant No.\,226-2024-00142). This work was partially carried out at the USTC Center for Micro and Nanoscale Research and Fabrication. The numerical calculations were performed
on the supercomputing system at the Supercomputing
Center of the University of Science and Technology of China.
\end{acknowledgments}

\bibliography{main}
\end{document}

\title{Supplementary Information to "Oxygen-vacancy quantum spin defects in silicon carbide"}
\maketitle

\tableofcontents % Generates the table of contents
\newpage

\section{Correlation Mapping of Stacking Fault and PL5/PL6} \label{secS: corr-map}
\subsection{Sample}
The sample used for correlation mapping of stacking faults and PL5/PL6 centers was diced from a wafer consisting of a 12.5-$\mu$m-thick intrinsic epitaxial layer of single-crystal 4H-SiC grown on a $4^\circ$ off-axis N-type 4H-SiC substrate. Registration markers were created on the epitaxial layer surface by laser ablation to facilitate repositioning. The sample was then implanted with 15-keV $^{14}\text{N}^{+}$ ions at a dose of approximately $1\times 10^{11} \ \text{cm}^{-2}$, followed by annealing at 1000\,$^\circ$C for 30 minutes in vacuum.
\begin{figure*}
 \includegraphics[scale=0.9]{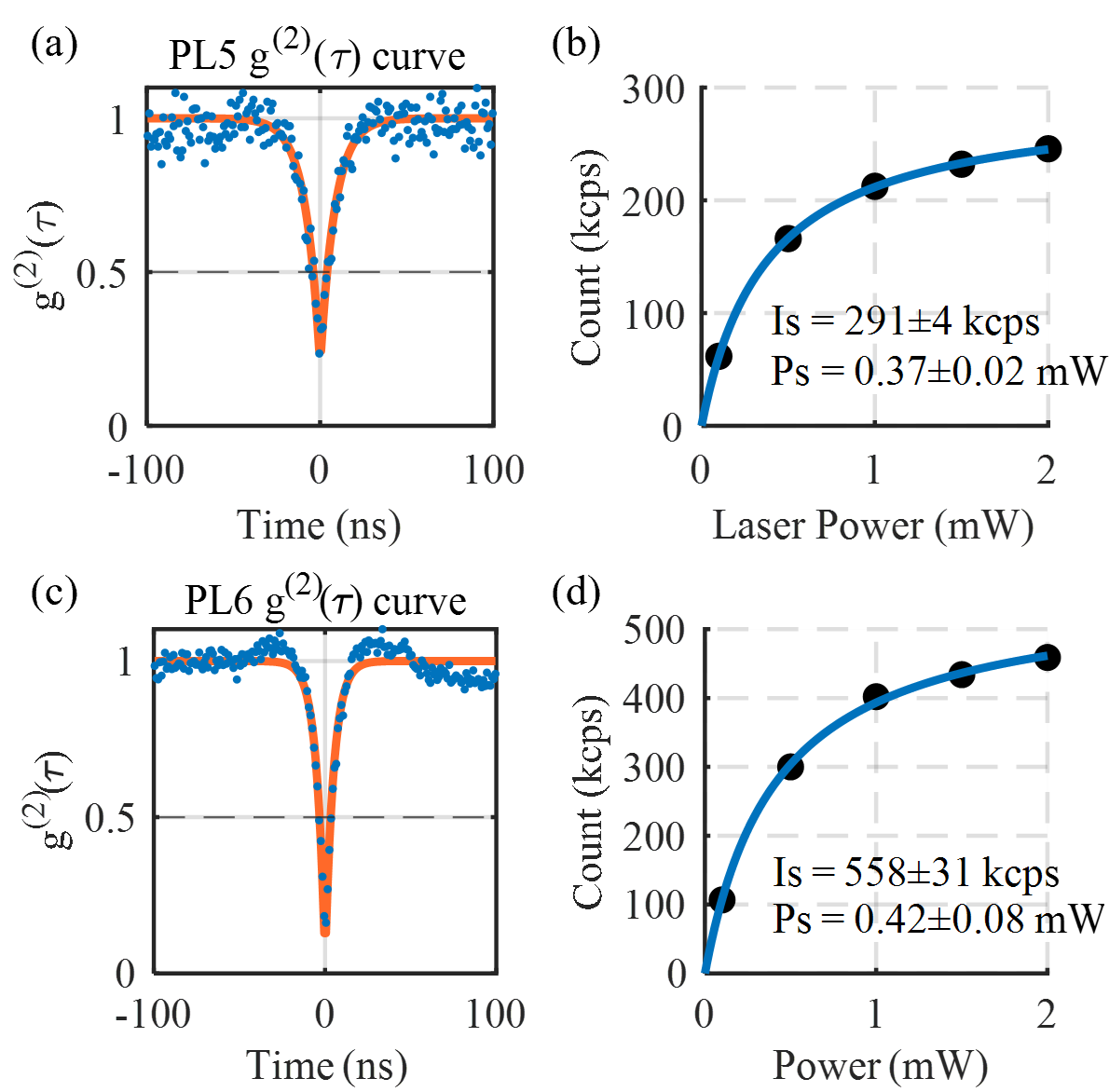}
 \caption{\label{figS: PL5-6 Saturation Count}(a) \(g^2(\tau)\) measurement of single PL5. Clearly the anti-bunching \(g^2(0)<0.5\) shows it's a single optical emitter. (b) Saturation behavior of the PL of singe PL5. The solid line was given by fitting I(P) = Is$\cdot$P/(P+Ps). (c) \(g^2(\tau)\) measurement of single PL6. (d) Saturation behavior of the PL of singe PL6. The solid line was given by fitting I(P) = Is$\cdot$P/(P+Ps). } 
 \end{figure*}

\subsection{Identification and localization of PL5 and PL6 centers}\label{subsecS:PL5-6Loc}

The locations of PL5 and PL6 centers were determined by first identifying all luminescent spots within the sample area and then screening for the specific spectroscopic signatures of PL5 and PL6.

\begin{figure*}
 \centering
 \includegraphics[scale=0.9]{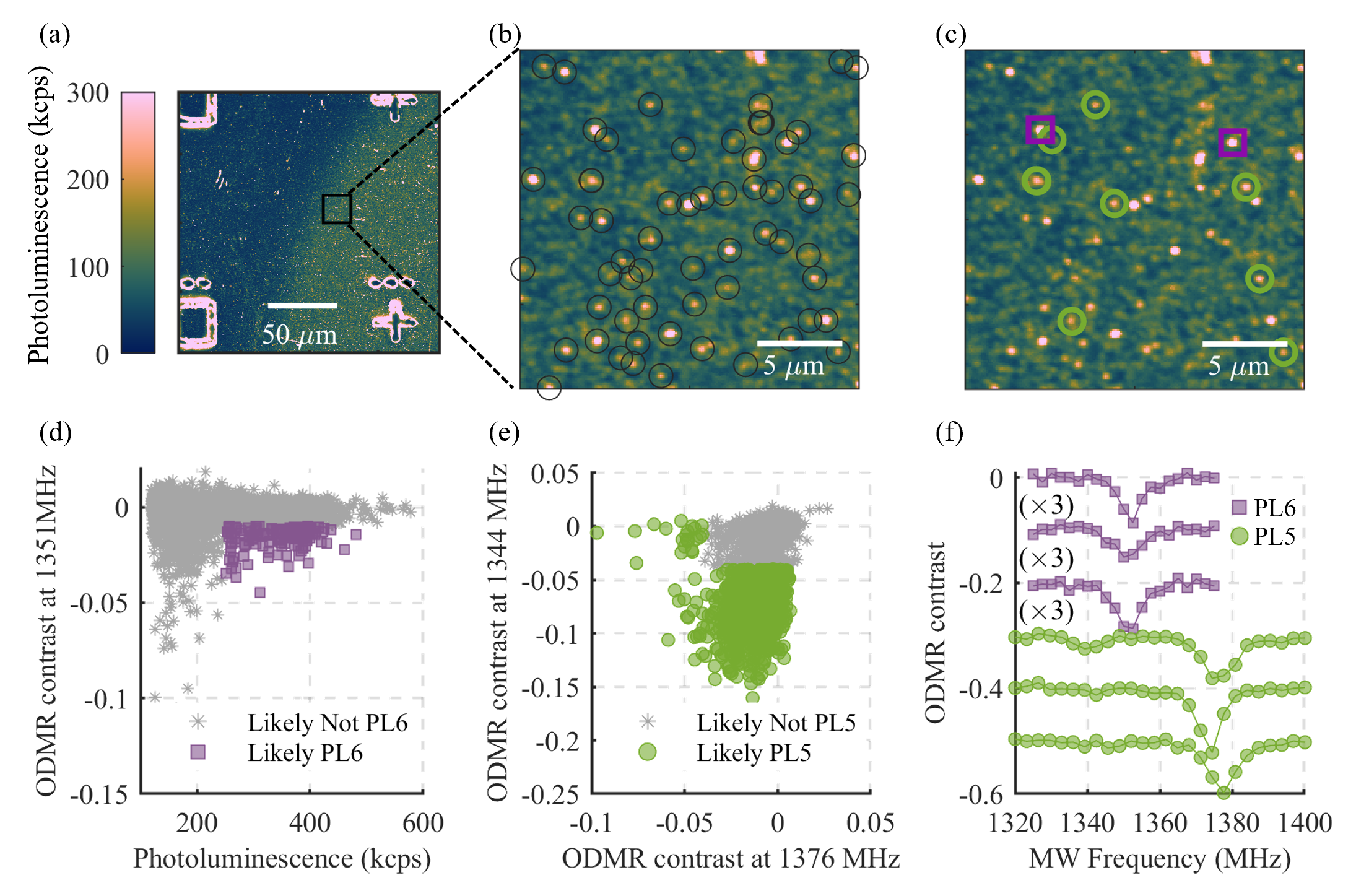}
 \caption{
 \textbf{(a)} Confocal image of the sample under 914~nm circularly polarized excitation (3~mW).
 \textbf{(b)} Magnified view of the region indicated in (a); spots exceeding 100~kcps are circled in black.
 \textbf{(c)} The same region as in (b), with confirmed PL5 centers circled in green and PL6 centers marked with purple squares.
 \textbf{(d)} Selection of candidate PL6 centers based on photon count $>250$~kcps and ODMR contrast at 1351~MHz $<-0.02$.
 \textbf{(e)} Selection of candidate PL5 centers using ODMR contrast thresholds at 1344~MHz or 1376~MHz $<-0.04$.
 \textbf{(f)} Representative zero‑field ODMR spectra of randomly chosen PL5 and PL6 centers from the final dataset.
 }
 \label{figS: CCFind_Image}
\end{figure*}

A confocal image of the sample region was acquired under excitation with a circularly polarized 914~nm laser (3~mW), as shown in \Cref{figS: CCFind_Image}\,(a). A magnified view [\Cref{figS: CCFind_Image}\,(b)] reveals numerous bright spots. Based on the saturation count rates of individual PL5 (180–300\,kcps) and PL6 (400--560\,kcps) under 0.4--0.8\,mW excitation (see \Cref{figS: PL5-6 Saturation Count}), a threshold of 100\,kcps was adopted for initial spot selection.

At zero magnetic field, PL6 exhibits an ODMR resonance near 1351~MHz, while PL5 shows two resonances near 1344 and 1376~MHz \cite{liRoomTemperatureCoherent2021}. These frequencies were used to distinguish PL5 and PL6 from the general pool of luminescent spots. Candidate PL6 centers were selected by requiring a photon count $>250$~kcps and an ODMR contrast at 1351~MHz below $-0.02$ [\Cref{figS: CCFind_Image}(d)]. Similarly, spots whose contrast at either 1344 or 1376~MHz fell below $-0.04$ were flagged as candidate PL5 centers [\Cref{figS: CCFind_Image}(e)].

Full ODMR spectra were then acquired for all candidate spots, allowing a final manual classification into PL5 and PL6. To illustrate the reliability of the procedure, \Cref{figS: CCFind_Image}\,(f) displays spectra of three randomly chosen PL5 and PL6 centers from the final dataset, showing the characteristic resonances near 1376~MHz for PL5 and 1351~MHz for PL6. The outcome of the screening applied to the region in \Cref{figS: CCFind_Image}(b) is presented in \Cref{figS: CCFind_Image}(c).

We note that the 1344~MHz resonance of PL5 is not visible in the spectra shown in \Cref{figS: CCFind_Image}(f) owing to the polarization characteristics of our microwave setup. This resonance requires linearly polarized excitation, whereas our radiating structure supplies a linear polarization nearly orthogonal to the required orientation. This was verified by applying a small magnetic field, which mixes the spin states and allows the resonance to be driven with circularly polarized microwaves, as confirmed on several PL5 centers.

\subsection{Stacking‑fault characterization by photoluminescence spectroscopy}

\begin{figure*}
 \centering
 \includegraphics[scale=0.9]{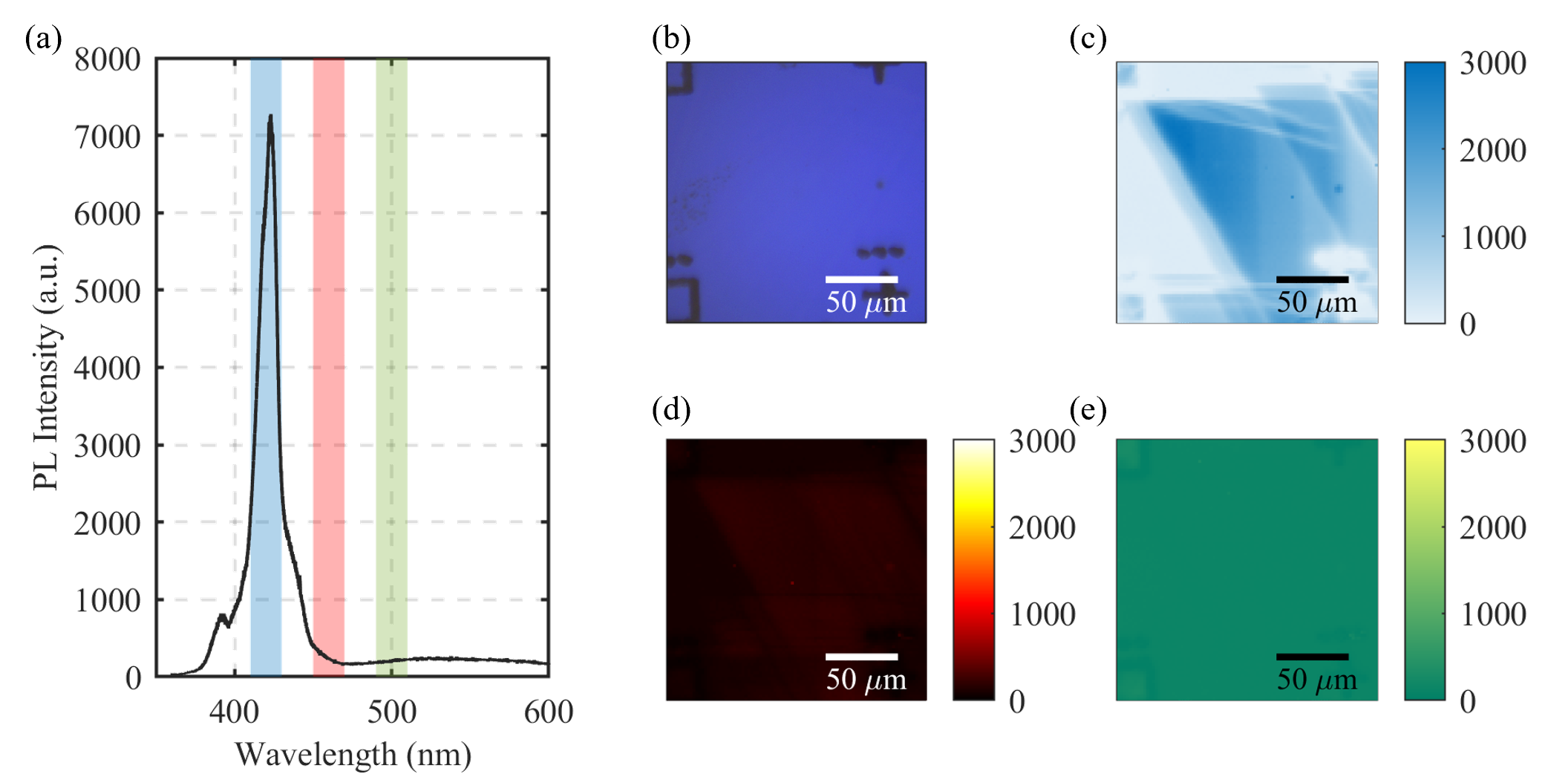}
 \caption{
 \textbf{(a)} PL spectrum collected from the central region of (b). The peak around 420~nm originates from a stacking fault, while the weaker peak near 390~nm corresponds to the band‑edge emission of 4H‑SiC \cite{fengCharacterizationStackingFaults2008}.
 \textbf{(b)} Optical image of the sample.
 \textbf{(c--e)} Maps of the integrated PL intensity in the ranges 410–430~nm (c), 450–470~nm (d), and 490–510~nm (e), each associated with a distinct type of stacking fault.
 }
 \label{figS: SF_PL_Image}
\end{figure*}
The stacking fault in the same region shown in \Cref{figS: CCFind_Image} was mapped using the photoluminescence (PL) method described in Ref.~\cite{fengCharacterizationStackingFaults2008}. Under 325~nm excitation, the intrinsic 4H-SiC PL spectrum exhibits a band-edge emission peak at 386~nm (3.21~eV). The presence of a stacking fault lowers the local band gap, resulting in a redshifted PL emission \cite{ivadyStabilizationPointdefectSpin2019, fengCharacterizationStackingFaults2008}, which enables its spatial mapping.
Room‑temperature photoluminescence (PL) mapping was performed using a LabRAM HR Evolution (HORIBA) system with 325~nm laser excitation. Spectra were collected from 350 to 600~nm [\Cref{figS: SF_PL_Image}(a)]. Three types of stacking faults in 4H‑SiC are known to produce PL bands centered near 420, 455, and 500~nm, respectively \cite{fengCharacterizationStackingFaults2008}. To map their spatial distribution, we integrated the PL intensity over the intervals 410–430~nm [\Cref{figS: SF_PL_Image}(c)], 450–470~nm [\Cref{figS: SF_PL_Image}(d)], and 490–510~nm [\Cref{figS: SF_PL_Image}(e)]. Only the stacking fault associated with the $\sim$420~nm emission was detected in our sample.

\subsection{Prediction of Stacking-fault-related model}
\begin{figure}[H] 
\centering 
\includegraphics{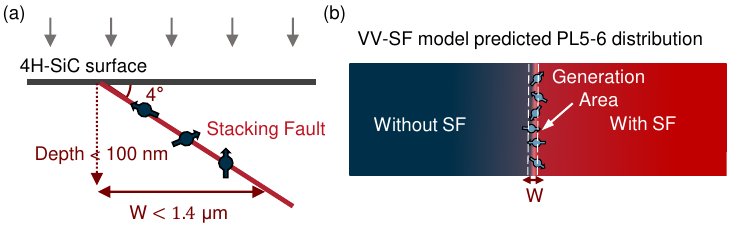}
\caption{(a) Schematic illustration of the VV-SF model. (b) Predicted distribution of PL5/PL6 color centers based on the VV-SF model.}
\label{figS:VV-SF_Prediction}
\end{figure}

According to the stacking-fault-related model, PL5/PL6 should be generated near the stacking fault, and the ion implantation depth further restricts the PL5/PL6 to be shallower than 100\,nm \cite{liNanoscaleDepthControl2019}. As the SF is inclined by $4^\circ$ relative to the sample surface in our sample, PL5/PL6 should be distributed along the intersection line between the SF and the surface, within an approximately \SI{1.4}{\micro\meter}-wide region, as illustrated in \Cref{figS:VV-SF_Prediction}(a-b).  However, contrary to the prediction of the stacking-fault-related model, PL5 and PL6 centers were observed at locations far from the stacking fault edge as shown in main text (Fig. 1). This spatial distribution demonstrates that the stacking-fault-related model does not apply to PL5 and PL6.

\newpage
\section{Spatial Mapping of PL5 and PL6 Centers in Oxygen-ion-implanted Samples}\label{secS: map-O-implanted}
\begin{figure*}
 \centering
 \includegraphics[]{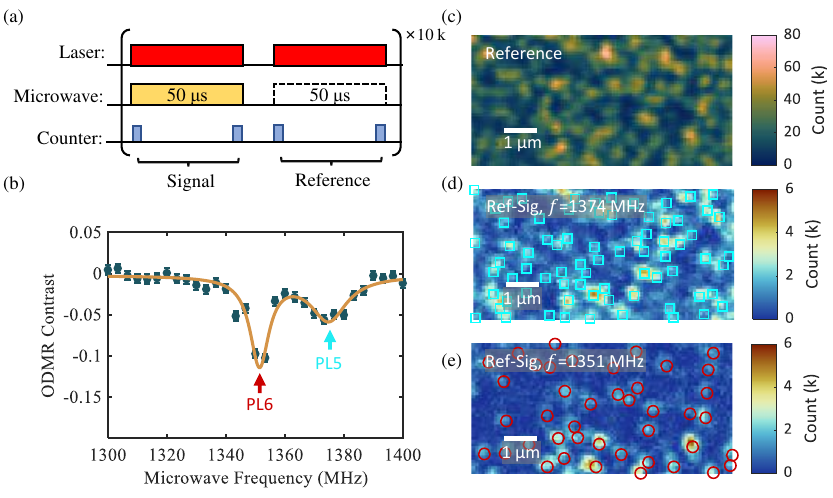}
 \caption{
 \textbf{(a)} Pulse sequence used for ODMR-based spatial mapping.
 \textbf{(b)} Full zero-field ODMR spectrum acquired in the oxygen-ion-implanted sample.
 \textbf{(c)} Reference photon-count map obtained with laser excitation only.
 \textbf{(d)} Difference map (reference $-$ signal) measured at $f = 1374~\text{MHz}$, highlighting PL5 centers.
 \textbf{(e)} Difference map measured at $f = 1351~\text{MHz}$, highlighting PL6 centers.
 }
 \label{figS: PL5-6 Loc of O-Implantation}
\end{figure*}

The spatial distribution of PL5 and PL6 centers in the oxygen-ion-implanted sample was mapped using the pulse sequence depicted in \Cref{figS: PL5-6 Loc of O-Implantation}(a). Reference photon counts ($I_{\text{ref}}$) were acquired under laser excitation alone, while signal counts ($I_{\text{signal}}$) were recorded with both laser excitation and resonant microwave irradiation. As shown in \Cref{figS: PL5-6 Loc of O-Implantation}(b), the optically detected magnetic resonance contrast, calculated as $(I_{\text{signal}} - I_{\text{ref}}) / I_{\text{ref}}$, varies with microwave frequency. PL5 exhibits a resonance at $f = 1374~\text{MHz}$, and PL6 resonates at $f = 1351~\text{MHz}$. Accordingly, at each pixel, the pulse sequence was applied at the respective resonant frequencies for PL5 and PL6, and both $I_{\text{ref}}$ and $I_{\text{signal}}$ were recorded. The resulting reference-count map is presented in \Cref{figS: PL5-6 Loc of O-Implantation}(c). Maps of the normalized signal difference $I_{\text{signal}} - I_{\text{ref}}$ at the PL5 and PL6 resonance frequencies are displayed in \Cref{figS: PL5-6 Loc of O-Implantation}(d) and (e), respectively. Based on these maps, the locations of PL5 and PL6 centers are indicated by circles and squares, respectively. The same procedure was applied to the nitrogen-ion-implanted sample discussed in the main text. The scanned areas and the numbers of identified PL5 and PL6 centers for both samples are summarized in \Cref{tabS: N-O Comparison}. All measurements were conducted under zero magnetic field.

\begin{table}[htbp]
\centering
\caption{Number of ODMR-detected PL5 and PL6 centers in samples implanted with oxygen or nitrogen ions under identical conditions (15~keV, $10^{11}~\text{cm}^{-2}$, annealed at 1050\,°C for 30~minutes).}
\begin{tabular}{ccccc}
\hline
 & \multicolumn{2}{c}{O implantation} & \multicolumn{2}{c}{N implantation} \\
\hline
Statistical area & \multicolumn{2}{c}{10$~\mu$m $\times$ 10$~\mu$m} & \multicolumn{2}{c}{10$~\mu$m $\times$ 10$~\mu$m} \\
Counting rate of single spot (cps) & \multicolumn{2}{c}{60--120\,k} & \multicolumn{2}{c}{$\sim$25\,k} \\
\hline
& PL5 & PL6 & PL5 & PL6  \\
\hline
Number of color centers  & $\ge$195 & $\ge$130 & $\sim$18 & $\sim$7 \\
ODMR contrast & 13\%--18\%& 8\%--14\%& 17\%--20\%& 16\%--20\%\\
Generation Efficiency & $\ge$1.95\textperthousand & $\ge$1.3\textperthousand & 0.18\textperthousand & 0.07\textperthousand \\
\hline
\end{tabular}
\label{tabS: N-O Comparison}
\end{table}

\newpage
\section{Orientation of PL5 Zero Field Splitting} 

\subsection{Orientation of axially symmetric spin splitting $D$}
\label{secS: orien-PL5}
In the main text, we have shown that PL5 exhibits six distinct orientations of the zero-field splitting parameter $D$, labeled D1 through D6. To determine these orientations accurately, we selected one representative color center for each of the six directions for further experimental characterization.

\begin{figure*}
 \includegraphics[scale=0.9]{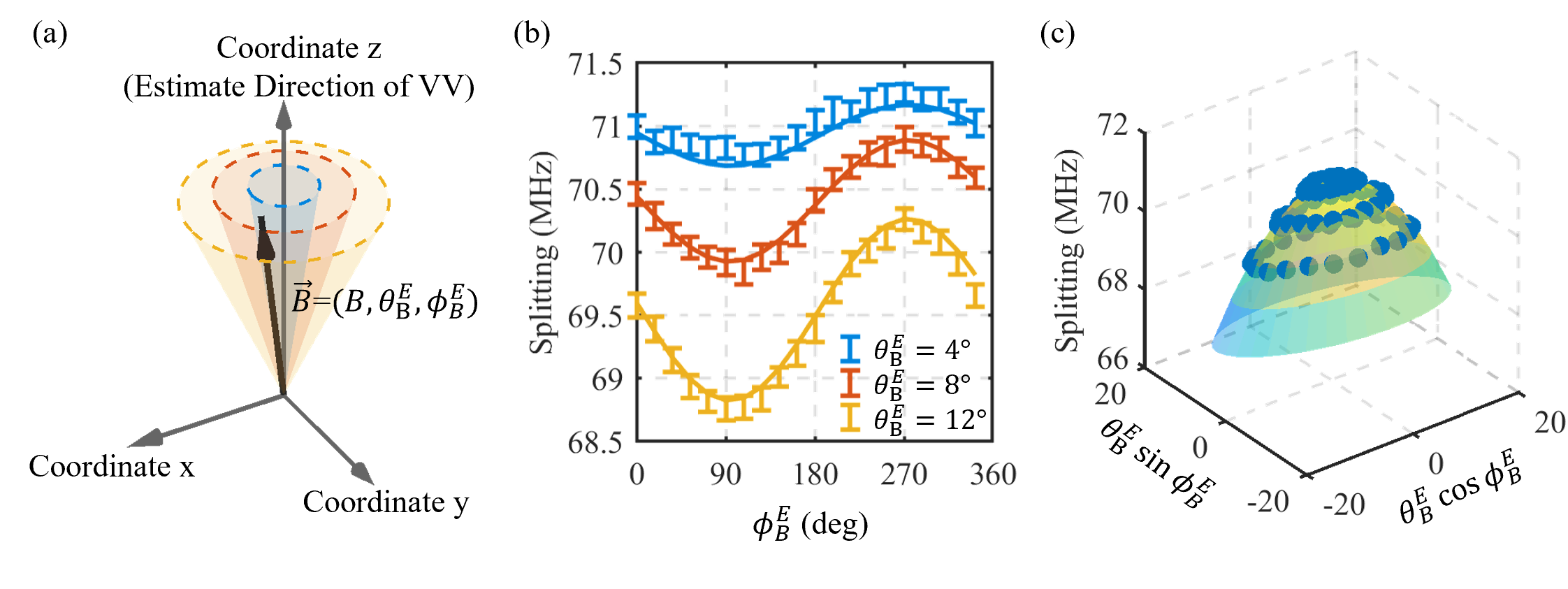}
 \caption{\label{Direction_Measurement_Fig}(a) Illustration of the direction measurement. In the experiment, we fixed the amplitude of $\vec{B}$ to be of $B= 25$ Gauss and let it scan in the spherical crown. For each magnetic field, we test the ODMR peak splitting. (b) ODMR peak splitting as a function of the direction $(\theta^E_B,\phi^E_B)$ of the applied magnetic field. The solid lines represent the results provided by the fitting.  (c) ODMR peak splitting as a function of the direction $(\theta^E_B,\phi^E_B)$ of the applied magnetic field. The surface represents the results obtained from the fitting. Fitting gives out the direction of D1 $\theta^E_{\text{D2}} = 2.8 \pm 0.2 \ \text{deg}$ and $\phi^E_{\text{D2}} = 86 \pm 3 \ \text{deg}$.} 
 \end{figure*}

For each center, the approximate orientation was initially known, and this direction was used to define the \textit{z}-axis of a local reference frame—referred to as the \textit{estimation coordinate system}—as illustrated in \Cref{Direction_Measurement_Fig}(a). A magnetic field \(\mathbf{B}\) , expressed in spherical coordinates \((B, \theta^E_B, \phi^E_B)\) within this estimation frame, was applied using a three-axis Helmholtz coil.
Let the orientation of the $D$ tensor of a given color center be represented in the same estimation coordinates as \((1,\theta^E_{\text{CC}}, \phi^E_{\text{CC}})\). The resulting ODMR splitting can then be described by 
\begin{equation}
    \text{Splitting} = 2 \gamma B (\cos{\theta^E_B} \cos{\theta^E_{\text{CC}}} + \sin{\theta^E_B} \sin{\theta^E_{\text{CC}}} \cos{(\phi^E_B-\phi^E_{\text{CC}})})
    \label{splitting_eq}
\end{equation}
By recording the splitting under various magnetic field orientations, we can fit the data using Eq.\,\eqref{splitting_eq} to extract the parameters \(\gamma B\), \(\theta^E_{\text{CC}}\), and \(\phi^E_{\text{CC}}\).
 \begin{table}[H] 
  \centering
  \setlength{\tabcolsep}{7mm}
  \begin{tabular}{cccc} %{\hsize}使三线表自适应宽度，c表示文本居中
    \toprule
    Label & $\gamma B $ (MHz) & $\theta^{\text{HC}}$ (deg)& $\phi^{\text{HC}}$ (deg)\\
    \midrule
PL5(D1)& 72.19$\pm$0.08& 108.4$\pm$0.5& 89.2$\pm$0.5\\
PL5(D2)& 71.18$\pm$0.04& 111.7$\pm$0.3& 28.1$\pm$0.3\\
PL5(D3)& 71.26$\pm$0.04& 111.3$\pm$0.3& -32.4$\pm$0.3\\
PL5(D4)& 70.55$\pm$0.08& 107.6$\pm$0.6& -93.8$\pm$0.6 \\
PL5(D5)& 71.16$\pm$0.08& 104.6$\pm$0.5& -153.9$\pm$0.5 \\
PL5(D6)& 73.01$\pm$0.05& 105.0$\pm$0.4& 148.8$\pm$0.4\\
 PL6(C-axis)& 137.87$\pm$0.05& 4.7$\pm$0.2&-5$\pm$3\\
     \bottomrule
  \end{tabular}
  \caption{Direction Measurement \Cref{splitting_eq} fitting results of different PL5-6 color centers. Results are given in Helmholtz Coil(HC) coordinate. We applied B=25 Gauss Magnetic field for PL5 and $B=50\ \text{G}$ for PL6.}
  \label{tabS: D_Direction}
\end{table}

In practice, we fixed the field magnitude at \(B = 25\ \text{G}\) and scanned the field direction around the initially estimated axis, as depicted in \Cref{Direction_Measurement_Fig}(a). For the D2 center, the measured ODMR splittings across different field orientations are shown in \Cref{Direction_Measurement_Fig}(b) and (c). The fitted orientation for this center is \(\theta^E_{\text{D2}} = 2.8 \pm 0.2^\circ\) and \(\phi^E_{\text{D2}} = 86 \pm 3^\circ\) in the estimation coordinate system.

Because each color center was analyzed in its own local estimation frame, we transformed all resulting orientations into a common reference frame—\textit{the Helmholtz coil (HC) coordinate system}—for consistent comparison. The complete set of fitted orientations for PL5 (D1--D6) is summarized in \Cref{tabS: D_Direction}.

\subsection{Orientation of transverse anisotropy splitting $E$} 
\label{secS: stress-orien-PL5}
We characterized the transverse zero-field splitting parameter $E$ for six individual PL5 centers, each exhibiting a distinct orientation (labeled D1 through D6). For each center, a coordinate system was carefully constructed such that the $z$-axis aligns with the orientation of $D$, and the $x$-axis lies in the plane formed by the $z$-axis and the $[0001]$ crystal direction. A magnetic field $\mathbf{B}$ was applied perpendicular to the $z$-axis ($\mathbf{B} \perp \hat{z}$).
Within this coordinate frame, the effective spin Hamiltonian for the PL5 center can be written as
\begin{equation}
    H / h = \frac{1}{\hbar^2} \left[ 
        D \hat{S}_z^2
        - E_x (S_x^2 - S_y^2) + E_y (S_x S_y + S_y S_x)
        + \gamma \hbar (B_x S_x + B_y S_y)
    \right],
\end{equation}
and the two magnetic dipole transition frequencies are given by \cite{dohertyTheoryGroundstateSpin2012,doldeElectricfieldSensingUsing2011}
\begin{equation}
    f_{\pm} \approx D + 3\eta \pm \left[ E - 2\eta \sin\theta \cos(2\varphi_B + \varphi_E) + \eta^2 \right]^{1/2},
    \label{eqS: fpm}
\end{equation}
where $\eta = B_{\perp}^2/(2D)$, $B_{\perp} = \sqrt{B_x^2 + B_y^2}$, $E = \sqrt{E_x^2 + E_y^2}$, and $\varphi_B$ and $\varphi_E$ denote the azimuthal angles of the in‑plane magnetic field and the transverse anisotropy vector $(E_x, E_y)$, respectively.

\begin{figure*}
 \includegraphics{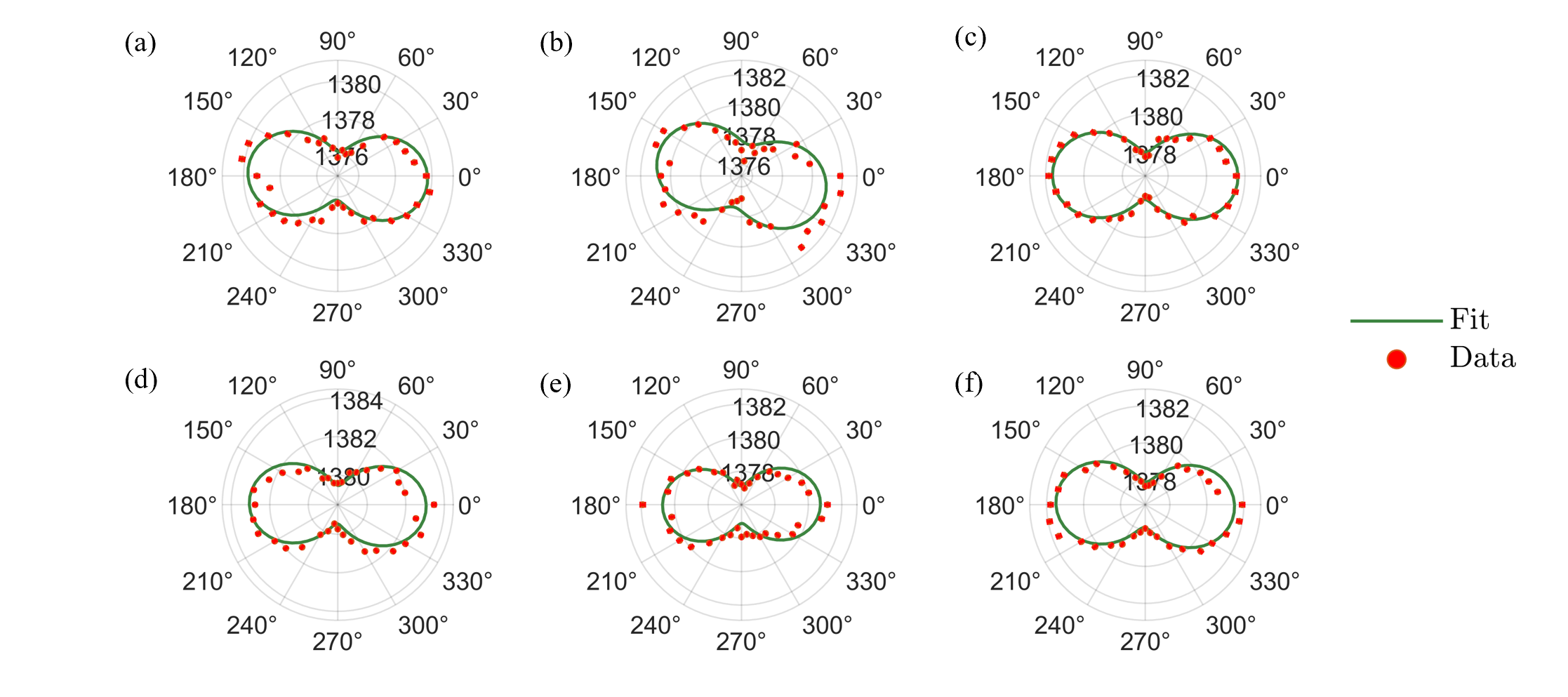}%
 \caption{\label{figS: D1-6 E Orientation}(a)$\sim$(f) Stress field measurements on six PL5 color centers in different $D$ orientations (from (a) to (f) corresponding to D2, D1, D6, D5, D4, D3). The angle of the polar plot denotes the azimuth angle $\varphi_B$ of the magnetic field. The amplitude of the polar plot denotes magnetic transition $f_+$. Red points represent the measured peak positions, while green lines show the fitting results.} 
\end{figure*}

Experimentally, the in‑plane field magnitude was fixed at $B_\perp = 25~\text{G}$, and the frequency $f_+$ was recorded as a function of the in‑plane field angle $\varphi_B$. The results for all six PL5 centers are presented in \Cref{figS: D1-6 E Orientation}. Solid lines in the figure correspond to fits using Eq.\,\eqref{eqS: fpm}, and the extracted parameters are summarized in \Cref{tabS: D1-6 E Orientation}.  Some deviations in the measurement results (like \Cref{figS: D1-6 E Orientation}(b)) may be attributed to systematic errors and the instability of certain color centers.
\begin{table}
  \caption{Fitting Results of direction for transverse anisotropy splitting $E$.}
  \centering
  \label{tabS: D1-6 E Orientation}
  \setlength{\tabcolsep}{5pt}
  \begin{tabular}{ccccccc}
    \toprule
    PL5 Orientation& D2& D1& D6& D5& D4&D3\\
    \midrule
    $\varphi_E$ (°)& $186.8 \pm 6.6$  & $207.2 \pm 10.4$& $182.2 \pm 3.5$& $183.7 \pm 6.0$& $178.2 \pm 6.7$&$184.6 \pm 5.1$\\
    \bottomrule
  \end{tabular}
\end{table}

\newpage
\section{Measurement of Hyperfine Couplings in PL5}
\label{secS:Measurement of Hyperfine Couplings in PL5}
\subsection{Sample}
The sample used for measurement of hyperfine couplings was diced from a wafer consisting of a 12.5-$\mu$m-thick intrinsic epitaxial layer of single-crystal 4H-SiC grown on a $4^\circ$ off-axis N-type 4H-SiC substrate. The sample was then implanted with 60-keV $^{14}\text{N}^{+}$ ions at a dose of approximately $1\times 10^{10} \ \text{cm}^{-2}$, followed by annealing at 1000\,$^\circ$C for 30 minutes in vacuum.

\subsection[Nearest C13 coupling]{Hyperfine coupling with $\mathrm{^{13}C_I}$ nuclei}
\begin{figure*}
    \centering
    \includegraphics[scale=0.9]{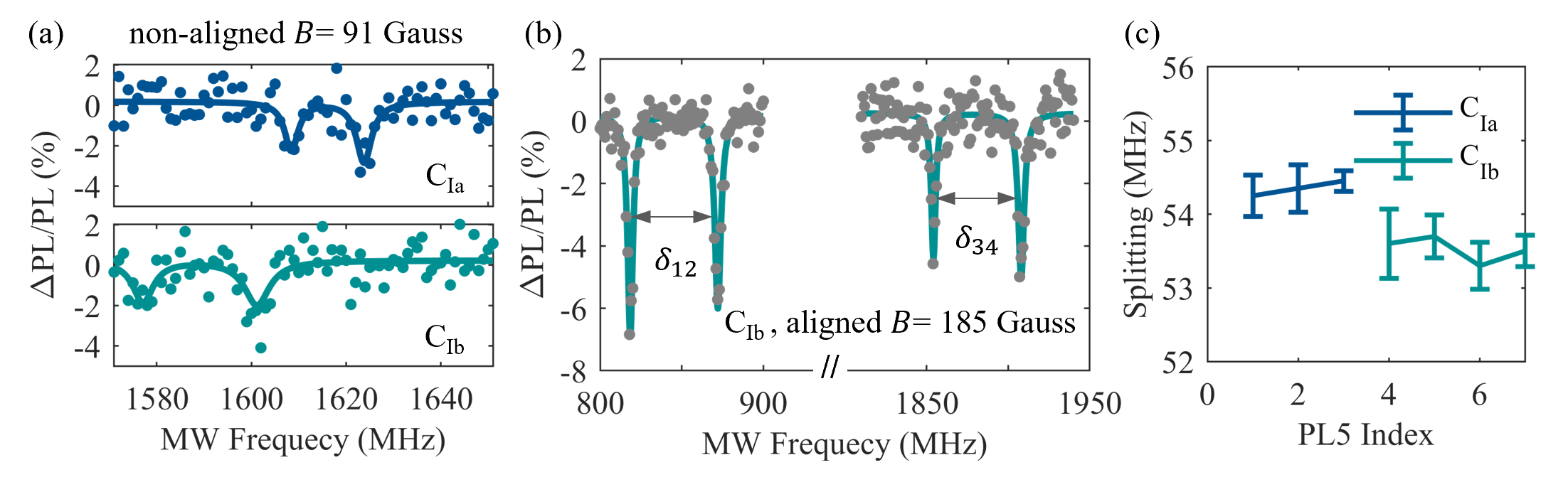}
    \caption{
        \textbf{(a)} ODMR spectra of a D2‑oriented PL5 center coupled to a $\text{C}_{\text{Ia}}$ (top) and a $\text{C}_{\text{Ib}}$ (bottom) nuclear spin, acquired with a magnetic field $B = 91\ \text{G}$ applied along the D5 direction. 
        \textbf{(b)} ODMR spectrum of a PL5 center coupled to a $\text{C}_{\text{Ia}}$ nuclear spin, measured under $B = 185\ \text{G}$. 
        \textbf{(c)} Mean hyperfine splitting $\delta = (\delta_{12}+\delta_{34})/2$ for $\text{C}_{\text{Ia}}$ and $\text{C}_{\text{Ib}}$ nuclear spins.}
    \label{figS: Nearest_C_Splitting}
\end{figure*}

In the experiment, the nuclear spin type (whether $\text{C}_{\text{Ia}}$ or $\text{C}_{\text{Ib}}$) can be identified from the ODMR spectrum obtained under a non‑aligned magnetic field [see \Cref{figS: Nearest_C_Splitting}(a)]. We measured the hyperfine splittings $\delta_{12}$ and $\delta_{34}$ under an axial magnetic field ranging from 90 to 200 G. As shown in \Cref{figS: Nearest_C_Splitting}(b), $\delta_{12}$ and $\delta_{34}$ exhibit a slight difference, which originates from the strong hyperfine coupling. 

While the individual splittings $\delta_{12}$ and $\delta_{34}$ are sensitive to both the magnitude and alignment accuracy of the applied field, their mean value $\delta = (\delta_{12}+\delta_{34})/2$ remains largely unaffected by these experimental variations. \Cref{figS: Nearest_C_Splitting}(c) plots $\delta$ for both $\text{C}_{\text{Ia}}$ and $\text{C}_{\text{Ib}}$ nuclear‑spin types; the corresponding numerical values are listed in \Cref{tabS: Nearest_C_Coupling}.
\begin{table}
    \centering
    \setlength{\tabcolsep}{10pt}
    \begin{tabular}{ccccc}
    \hline\hline
         PL5 Direction&  Nuclear Location&  B (Gauss)&  $\delta_{12}$ (MHz)&  $\delta_{34}$ (MHz)\\
         \hline
 D5& Unclassified& 91& 54.4$\pm$1.3&53.7$\pm$2.1\\
         D5&  $\text{C}_{\text{Ia}}$&  91&  54.2$\pm$0.5&  53$\pm$0.8\\
         D5&  $\text{C}_{\text{Ib}}$&  91&  53.9$\pm$0.4&  54.6$\pm$0.4\\
         D2&  $\text{C}_{\text{Ib}}$&  185&  55.3$\pm$0.5&  53.4$\pm$0.4\\
         D2&  $\text{C}_{\text{Ia}}$&  185&  53.7$\pm$0.3&  53.7$\pm$0.5\\
 D4& $\text{C}_{\text{Ib}}$& 185& 54.0$\pm$0.2& 54.9$\pm$0.2\\
 D1& $\text{C}_{\text{Ia}}$& 168& 52.0$\pm$0.4& 54.6$\pm$0.5\\
 D1& $\text{C}_{\text{Ia}}$& 168& 54.3$\pm$0.3& 52.7$\pm$0.3\\
 \hline\hline
    \end{tabular}
    \caption{Measured splitting $\delta_{12}$ and $\delta_{34}$ of different PL5. The uncertainty is given at a 95\% confidence level.}
    \label{tabS: Nearest_C_Coupling}
\end{table}
\subsection{Hyperfine coupling with other nuclear spins}
\label{secS:Hyperfine Coupling with Other Nuclear Spins}
\subsubsection{method}\label{secS: HF_method}
When a single nuclear spin couples to the PL5 center, the Hamiltonian of the PL5 ground state can be written as \cite{miaoElectricallyDrivenOptical2019,dohertyTheoryGroundstateSpin2012}

\begin{equation}
\begin{gathered}
H / h =\frac{1}{\hbar^2} \Bigl( D \hat{S}_z^2
-  E_x(S_x^2-S_y^2) +  E_y(S_x S_y + S_y S_x) \\
+ \gamma \hbar \mathbf{B} \cdot  \hat{\mathbf{S}}
+ \hat{\mathbf{S}} \cdot \mathbf{A} \cdot \hat{\mathbf{I}} 
+ \gamma_\mathrm{C} \hbar \mathbf{B} \cdot  \hat{\mathbf{I}}  \Bigr),
\end{gathered}
\end{equation}
where $D$ is the longitudinal zero-field splitting, $E_x$ and $E_y$ are the transverse zero-field splitting components, $\gamma$ denotes the electron gyromagnetic ratio, $\hat{\mathbf{S}}$ and $\hat{\mathbf{I}}$ are the electron and nuclear spin operators, respectively, $\mathbf{A}$ is the hyperfine tensor for a spin‑1/2 nucleus, and $\mathbf{B}$ is the external magnetic field. The nuclear Zeeman term is neglected in the following. We adopt a coordinate system in which the $z$‑axis coincides with the orientation of the PL5 center, and the $x$‑axis lies in the symmetry plane of the 4H‑SiC crystal and is perpendicular to $z$.

In the regime $\mathbf{B},\mathbf{A} \ll D$, first-order perturbation theory yields four transition frequencies
\begin{equation}
\label{Eq-f}
    f \approx D \pm \sqrt{(\gamma B_z\pm A_z/2)^2+E^2},
\end{equation}
with $E=\sqrt{E_x^2 + E_y^2}$ and $A_z = | \hat{z} \cdot \mathbf{A} |$. These frequencies are labeled $f_{1,2,3,4}$ in increasing order.
\begin{figure*}
 \centering
 \includegraphics[scale=0.9]{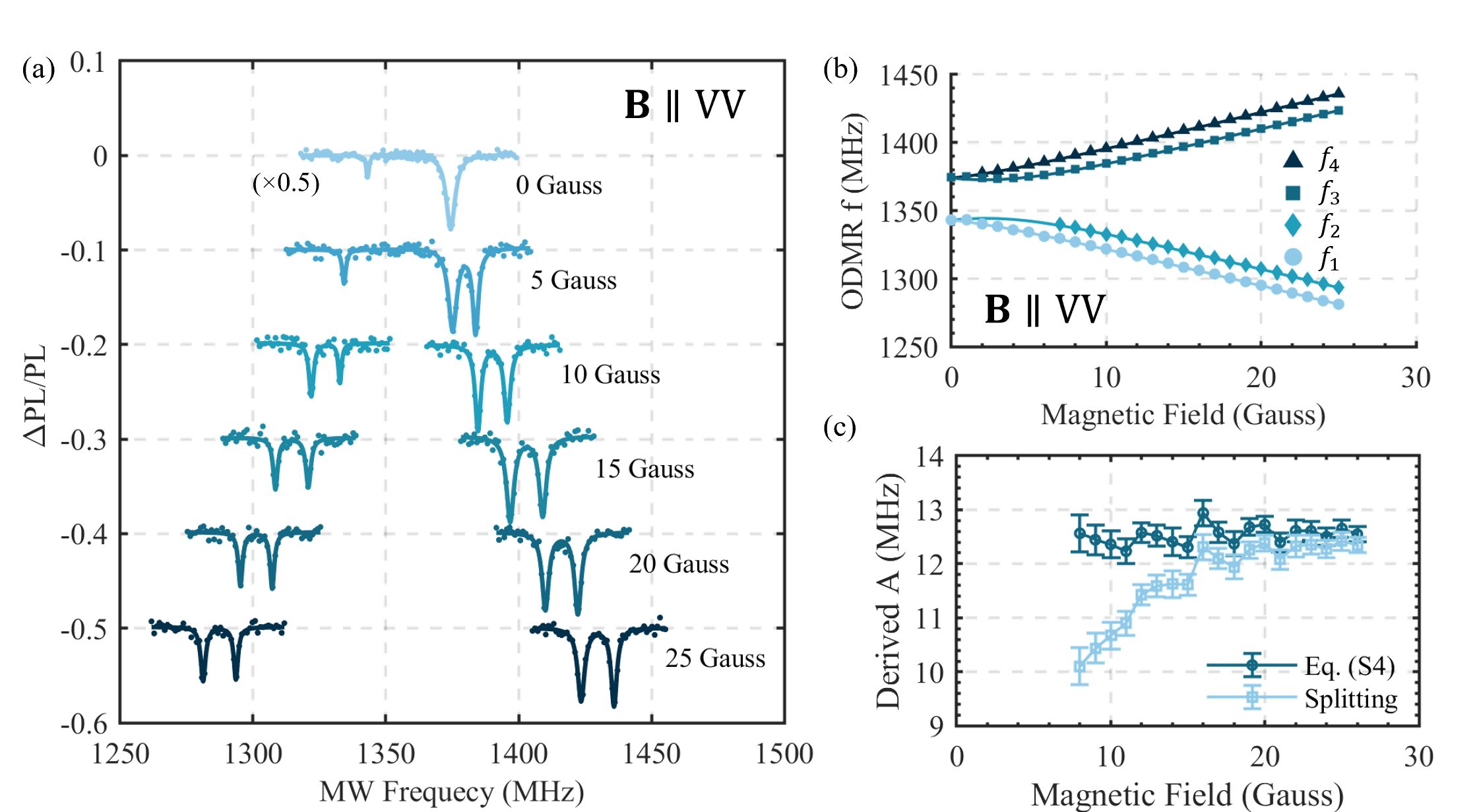}
 \caption{
 \textbf{(a)} ODMR spectra of a selected PL5 center under different magnetic fields. The zero-field spectrum is scaled down by a factor of 2 for clarity. 
 \textbf{(b)} Measured transition frequencies $f_{1,2,3,4}$ (points) as a function of magnetic field. Solid curves are calculated from Eq.\,\eqref{Eq-f} with $D=1358.5\ \text{MHz}$, $E=14.1\ \text{MHz}$, and $A_z=12.5\ \text{MHz}$.
 \textbf{(c)} Hyperfine component $A_z$ derived via Eq.\,\eqref{Eq-A} (dark green) and via the direct splitting $A_z^{\text{(split)}}$ (light blue).}
 \label{figS: hyperfine-method}
\end{figure*}

As illustrated in \Cref{figS: hyperfine-method}(a), a single PL5 coupled to a single nuclear spin shows no resolvable hyperfine splitting at zero magnetic field because of the transverse zero-field term. When a finite magnetic field is applied, the splitting appears and eventually saturates. Experimentally measured transition frequencies as a function of field are presented in \Cref{figS: hyperfine-method}(b); they are well described by Eq.\,\eqref{Eq-f} with appropriate parameters.

In the main text, the hyperfine component $A_z$ is extracted via
\begin{equation}
    \label{Eq-A}
    A_z = \frac{(f_4-f_1)^2-(f_2-f_3)^2}{8 \gamma B_z},
\end{equation}
which follows directly from Eq.\,\eqref{Eq-f}. Here $f_{1,2,3,4}$ are obtained from the ODMR spectra, and $\gamma B_z$ must be determined independently. We adopt $\gamma = 2.8\ \text{MHz/G}$ and assume that the magnetic field is well aligned with the PL5 axis, so that $B_z \approx B$, where $B$ is the nominal field set by the Helmholtz coils. The small discrepancy between the nominal and actual field (discussed in later sections) is negligible for the hyperfine determination. To verify this assumption, we fit the quantity $(f_4-f_1)^2 + (f_3-f_2)^2$ to $8aB^2+b$, yielding $\gamma B_z = \sqrt{a}\,B$. The fit gives $\gamma B_z = 2.786(4)\ \text{MHz/G} \times B$, in excellent agreement with the assumed value.

\Cref{figS: hyperfine-method}(c) compares the hyperfine component $A_z$ computed from Eq.\,\eqref{Eq-A} with the simple hyperfine splitting $A_z^{\text{(split)}} = [(f_2-f_1)-(f_4-f_3)]/2$, using the same data as in \Cref{figS: hyperfine-method}(b). Eq.\,\eqref{Eq-A} clearly provides a more consistent result across different magnetic fields.

\subsubsection{hyperfine statistics}
\label{secS: hyper_stati}
To survey the hyperfine properties of PL5 centers, we performed ODMR spectroscopy on 473 individual PL5 centers in an off‑axis magnetic field of $15\ \text{G}$. 

The six possible orientations of PL5 form an angle of about $71^\circ$ with the crystal $c$‑axis. For the hyperfine survey, we divided them into three groups (illustrated in \Cref{figS: Bz-Test}(a–c)). For each group we aimed to align the magnetic field so that its angle relative to every center in the group was approximately equal. In practice, small misalignments were unavoidable; their effect is quantified in the next subsection. The values of $\gamma B_z$ used for the hyperfine analysis in each group are listed in \Cref{tabS: hyper_Bz}.

\begin{figure*}
 \centering
 \includegraphics[scale=0.9]{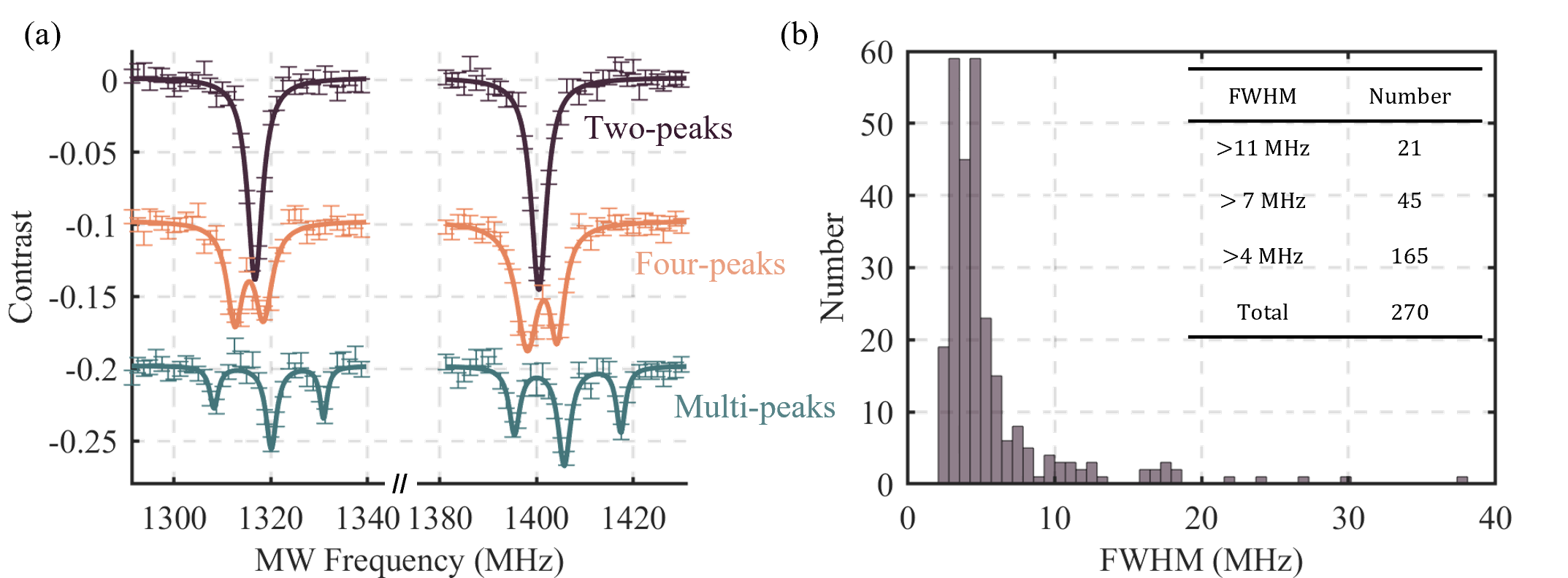}
 \caption{
 \textbf{(a)} Representative ODMR spectra acquired under a $15\ \text{G}$ off‑axis magnetic field. 
 \textbf{(b)} Distribution of the full‑width‑at‑half‑maximum (FWHM) for the two‑peak centers shown in (a).}
 \label{figS: B15_OA_ODMR}
\end{figure*}

Representative spectra are displayed in \Cref{figS: B15_OA_ODMR}(a). According to the number of resolved resonance peaks, we classified the centers into three categories: (1) two peaks (no strong hyperfine coupling), (2) four peaks (coupling to a single nuclear spin), and (3) multiple peaks (coupling to several nuclear spins). The classification results are summarized in \Cref{tabS: ClassifyT}. Note that centers with weak hyperfine coupling may be misclassified as “no coupling” if the splitting is obscured by line broadening. To assess this possibility, we analyzed the distribution of ODMR linewidths for the two‑peak centers, as shown in \Cref{figS: B15_OA_ODMR}(b).

\begin{table}[H]
\centering
\caption{Classification of PL5 ODMR spectra acquired at an applied magnetic field of $B = 15\ \text{G}$. Among the spectra that show coupling to single nuclear spin, there are 11 of them corresponding to hyperfine interaction with $\mathrm{C_I}$ nuclei.}
\label{tabS: ClassifyT}
\begin{tabular}{ccc}
\toprule
Coupling type & Number & Percentage \\
\midrule
No nuclear spin & 276 & 58.4\% \\
Single nuclear spin & 139 & 29.4\% \\
Multiple nuclear spins & 34 & 7.2\% \\
Unresolved & 24 & 5.1\% \\
\hline
Total & 473 & 100\% \\
\bottomrule
\end{tabular}
\end{table}

\subsubsection[Extent of $\gamma B_z$ Deviation]{extent of the $\gamma B_z$ deviation in hyperfine calculations}
\label{secS: Bz_deviation}
For the hyperfine survey we applied a field $B = 15\ \text{G}$ at an intended angle of $19.5^\circ$ relative to the PL5 axis, giving an expected $\gamma B_z \approx 2.8 \times 15 \times \cos 19.5^\circ = 39.6\ \text{MHz}$. Deviations from this value arise from two main sources: (1) the finite setting accuracy of the Helmholtz coils, which introduces an uncertainty of about $1\ \text{MHz}$ at $15\ \text{G}$ (see \Cref{secS: HC-cali}); and (2) small angular misalignments between the intended field direction and the actual PL5 orientation, quantified in \Cref{tabS: hyper_Bz}. For each group we mitigate the latter by using the average $\gamma B_z$ of the two orientations in that group. As seen from the table, the angular contribution to the uncertainty is also about $1\ \text{MHz}$. Combining both sources, we estimate the total uncertainty in $\gamma B_z$ to be $\sim 1\ \text{MHz}$ for all groups.

\begin{figure*}
 \centering
 \includegraphics[scale=0.9]{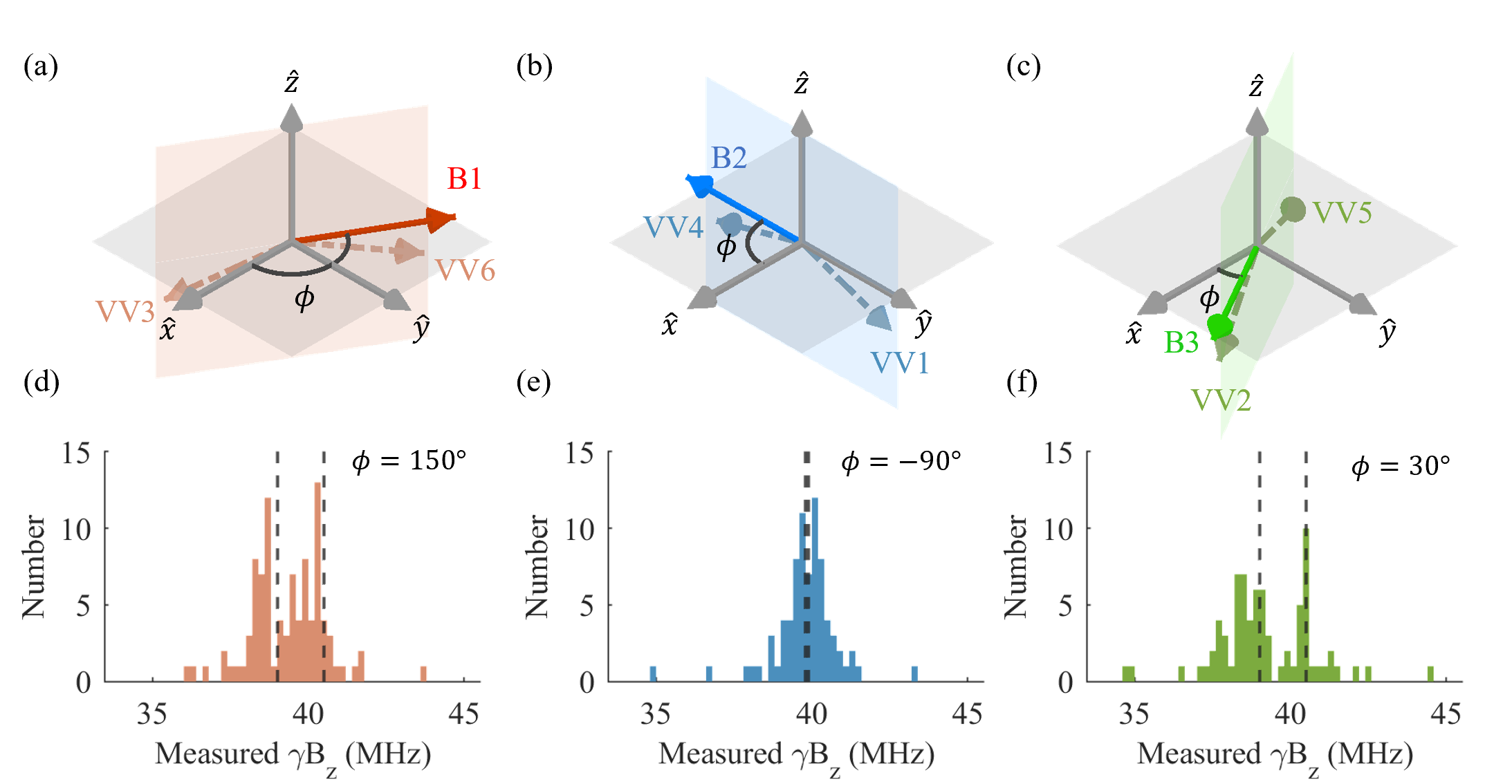}
 \caption{
 \textbf{(a–c)} Schematic of the magnetic‑field alignment for the three orientation groups defined in \Cref{tabS: hyper_Bz}. 
 \textbf{(d–f)} Distributions of $\gamma B_z$ extracted from two‑peak centers via Eq.\,\eqref{eqS: Bz-two-peaks}. Dashed lines mark the $\gamma B_z$ values in \Cref{tabS: hyper_Bz}.}
 \label{figS: Bz-Test}
\end{figure*}

We cross‑checked this estimate by analyzing the PL5 color centers identified as no strong hyperfine coupling in the hyperfine survey (\Cref{tabS: ClassifyT}). For a PL5 without hyperfine coupling, the axial field component can be obtained from the two observed transition frequencies via
\begin{equation}
\label{eqS: Bz-two-peaks}
    \gamma B_z = \sqrt{\bigl[(f_2-f_1)/2\bigr]^2-E^2}
\end{equation}
where $f_{1,2}=D\mp\sqrt{(\gamma B_z)^2+E^2}$. The distributions of $\gamma B_z$ obtained from Eq.\,\eqref{eqS: Bz-two-peaks} for the three groups are shown in \Cref{figS: Bz-Test}(d–f). Dashed lines indicate the $\gamma B_z$ values in \Cref{tabS: hyper_Bz}; the agreement is good apart from a few outliers.
\begin{table}[]
\setlength{\tabcolsep}{5pt}
\centering
\caption{Magnetic‑field components $\gamma B_z$ for the three orientation groups. The PL5 orientations $\theta^{\text{HC}}_{\text{CC}},\phi^{\text{HC}}_{\text{CC}}$ are taken from \Cref{tabS: D_Direction}. The gyromagnetic ratio is $\gamma = 2.8\ \text{MHz/G}$. The last column is $\gamma B_z$ used in hyperfine $A_z$ calculation.}
\label{tabS: hyper_Bz}
\begin{tabular}{cccccc}
\toprule
\makecell[c]{Field direction\\(HC coordinates)} & PL5 & $\theta_{\text{CC}}$ (deg) & $\phi_{\text{CC}}$ (deg) & \makecell[c]{$\gamma B_z$ (MHz)} & \makecell[c]{$\gamma B_z$ (MHz)\\(used for $A_z$ calculation)} \\
\midrule
\multirow{2}{*}{\makecell[c]{$\theta_{\text{B}}=90^\circ$\\$\phi_{\text{B}}=-90^\circ$}} 
 & D1 & $108.4\pm0.5$  & $89.2\pm0.5$ & 39.85 & \multirow{2}{*}{39.90} \\
 & D4 & $107.6\pm0.6$  & $-93.8\pm0.6$ & 39.95 \\
\midrule
\multirow{2}{*}{\makecell[c]{$\theta_{\text{B}}=90^\circ$\\$\phi_{\text{B}}=30^\circ$}} 
 & D2 & $111.7\pm0.3$  & $28.1\pm0.3$ & 39.00 & \multirow{2}{*}{39.76} \\
 & D5 & $104.6\pm0.5$  & $-153.9\pm0.5$ & 40.55 \\
\midrule
\multirow{2}{*}{\makecell[c]{$\theta_{\text{B}}=90^\circ$\\$\phi_{\text{B}}=150^\circ$}}  
 & D3 & $111.3\pm0.3$  & $-32.4\pm0.3$ & 39.10 & \multirow{2}{*}{39.83} \\
 & D6 & $105.0\pm0.4$  & $148.8\pm0.4$ & 40.56 \\
\bottomrule
\end{tabular}
\end{table}
\newpage

\section{ Evaluate the Accuracy of Helmholtz's Magnetic Field Setting} 
\label{secS: HC-cali}
\begin{figure}[H]
\centering
 \includegraphics{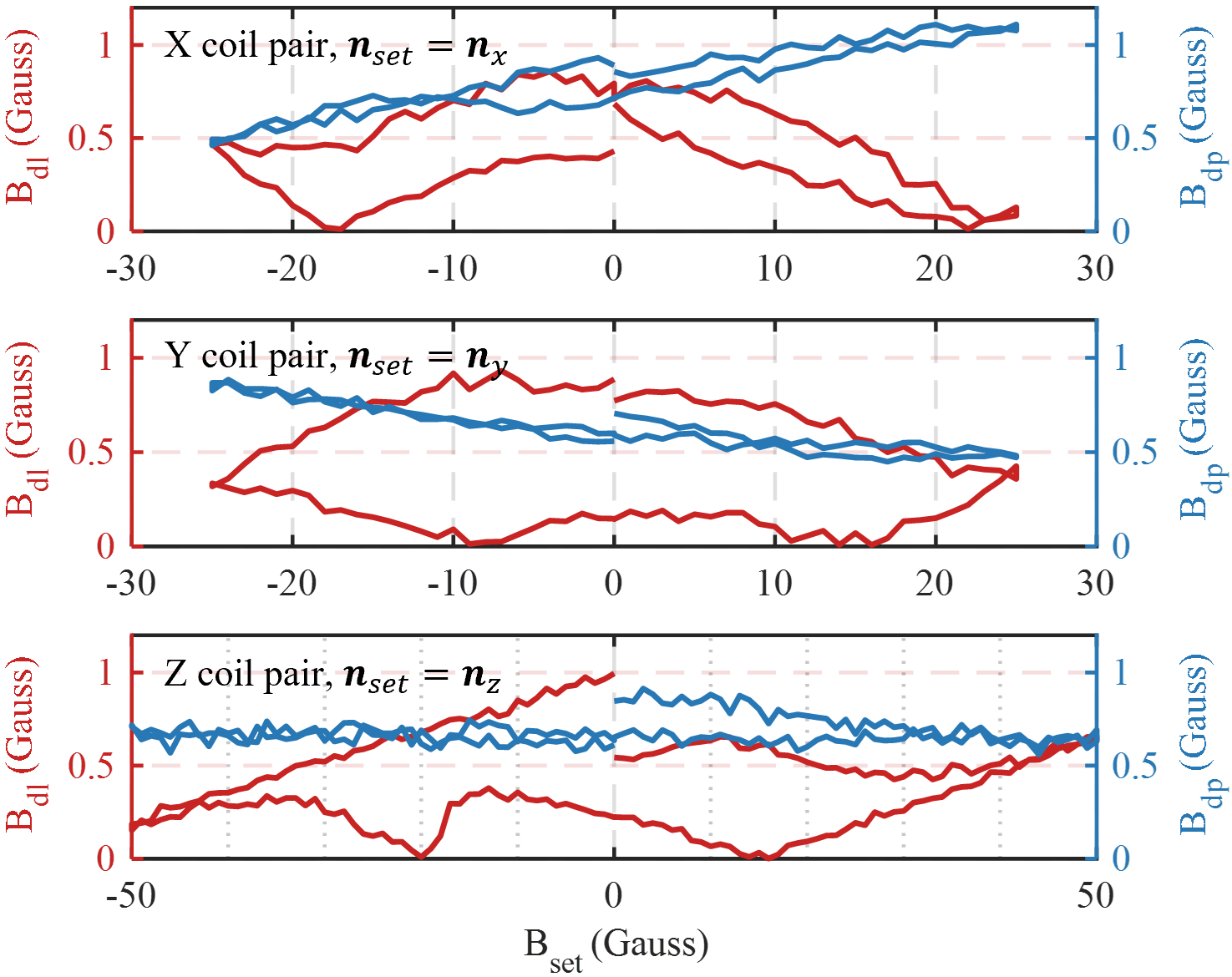}
\caption{Magnetic field deviations \(\mathbf{B}_\mathrm{d}=\mathbf{B}_\mathrm{dl}+\mathbf{B}_\mathrm{dp}\), measured as a function of the set field \(\mathbf{B}_{\text{set}}=\text{B}_{\text{set}}\mathbf{n}_{\text{set}}\), where \(\mathbf{B}_\mathrm{dl}\) is parallel to \(\mathbf{B}_{\text{set}}\) and \(\mathbf{B}_\mathrm{dp}\) is perpendicular to \(\mathbf{B}_{set}\). The field \(\text{B}_{\text{set}}\) was initially set to zero, then gradually increased, decreased to a negative value, and finally returned to zero. The primary cause of the deviation is the magnetism of the objective.}
\label{figS: HC-cali}
 \end{figure}

COLIY G93 gaussmeter was used for the calibration of the three-axis Helmholtz coils. Our three-axis Helmholtz coils are composed of three pairs of Helmholtz coils, namely X, Y, and Z coils. For each pair of coils, we set the magnetic field to be $\mathbf{B}_{\text{set}}=\text{B}_{\text{set}}\mathbf{n}_{\text{set}}$  and measure the actual magnetic field $\mathbf{B}_{\text{mea}}$. The deviation is given by $\mathbf{B}_\mathrm{d} = \mathbf{B}_{\text{mea}} - \mathbf{B}_{\text{set}}$.  We further divide \(\mathbf{B}_\mathrm{d}\) as  \(\mathbf{B}_\mathrm{d}=\mathbf{B}_\mathrm{dl}+\mathbf{B}_\mathrm{dp}\) in which \(\mathbf{B}_\mathrm{dl}//\mathbf{B}_{\text{set}}\) and \(\mathbf{B}_\mathrm{dp}\perp \mathbf{B}_{\text{set}}\). 
The measured deviation for \(\mathbf{B}_\mathrm{dl}\) and $\mathbf{B}_\mathrm{dp}$ are both $\sim$ 1 Gauss which is mainly due to the magnetism of the objective (\Cref{figS: HC-cali}).
\newpage
\section{First-principles Computational Details}

First-principles calculations were performed to investigate the electronic structure and spin properties of neutral oxygen-substituted silicon vacancies and divacancies using density functional theory (DFT). All calculations were carried out using the Vienna \textit{ab initio} Simulation Package (VASP)\,\cite{kresse1993ab, kresse1994ab} employing the projector augmented wave (PAW) method\,\cite{blchl1994}. The structural relaxations utilized the Revised Perdew-Burke-Ernzerhof functional for solids (PBEsol)\,\cite{PhysRevLett.100.136406}. The plane-wave cutoff energy was set to \SI{520}{\electronvolt}. A $12\times 12 \times 4$ Monkhorst-Pack k-point mesh was used for the primitive cell, while a single $\Gamma$ point was employed for the expanded supercell calculations. The convergence criteria were set to \SI{1e-6}{\electronvolt} for the self-consistent field (SCF) energy and \SI{0.01}{\electronvolt\per\angstrom} for the ionic relaxation forces. To simulate the defects, we constructed a quasi-cubic 478-atom supercell defined by the basis vectors $3\bm{a}+6\bm{b}$, $5\bm{a}$, and $2\bm{c}$, where $(\bm{a}, \bm{b}, \bm{c})$ represent the primitive translation vectors of 4H-SiC. The resulting supercell dimensions are $16.07$, $15.48$, and $20.25\ \text{\AA}$. To ensure the identification of the global ground state, symmetry constraints were lifted during relaxation, and small random perturbations were applied to the initial atomic coordinates. Consequently, the optimized structures may exhibit minor deviations from ideal symmetry.

The zero-phonon line (ZPL) energies and charge transition levels were estimated based on single-particle Kohn-Sham orbitals at the equilibrium geometry of the spin-triplet states. The zero-field splitting (ZFS) parameters were calculated using the formalism based on periodic density functional calculations described in Ref.\,\cite{PhysRevB.77.035119}. 

For the hyperfine coupling (HFC) calculations, the HSE06 hybrid functional was employed to improve the description of the electronic localization and spin density\,\cite{10.1063/1.2404663}. The hyperfine tensors were calculated taking into account core electronic contributions using the frozen valence approximation\,\cite{yazyev2005core}. The nuclear gyromagnetic ratios for $\rm ^{29}$Si and $\rm ^{13}$C were taken as $-8.465$\,MHz/T and $10.705$\,MHz/T, respectively. 

\begin{figure}[H]
\centering
 \includegraphics[scale=1.3]{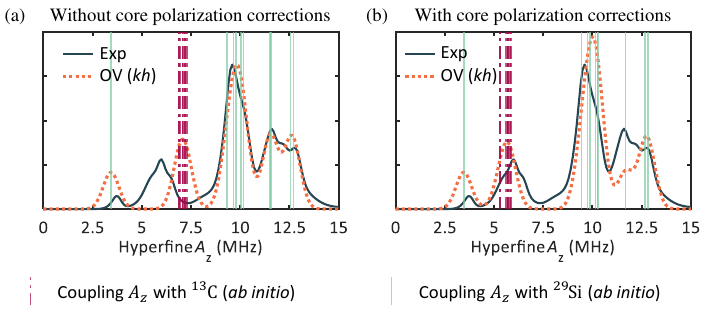}
\caption{Comparison of experimental (black solid curve) and simulated (orange dashed curve) $A_z$ distributions illustrating the impact of core polarization. (a) Simulated hyperfine distribution without core polarization corrections. (b) Simulated hyperfine distribution with core polarization corrections included.}
\label{figS: Hyper_CorePolari_compare}
\end{figure}

To rigorously evaluate the impact of core polarization on the HFC results, we compared the computed hyperfine distributions with and without core electronic contributions for the $\mathrm{O_C V_{Si}}(kh)$ defect, as shown in \Cref{figS: Hyper_CorePolari_compare}. The inclusion of core polarization leads to measurable shifts in the peak positions for $\mathrm{^{13}C}$ nuclei [compare \Cref{figS: Hyper_CorePolari_compare}(a) and (b)], although the relative spectral line shapes and overall distributions remain largely unaffected. In the main text, we present results including core polarization corrections to provide the most physically accurate description. The uncorrected results presented here serve as a reference for comparison with prior theoretical studies that utilized standard setups (e.g., Ref.\,\cite{zhaoPL6Calc2025}). The consistency of the qualitative conclusions under both computational schemes further confirms the robustness of the derived hyperfine characteristics.

.

\newpage
%\bibliographystyle{unsrt}
\bibliographystyle{apsrev4-2}
\bibliography{SupBibtex.bib}